\documentclass[twocolumn]{aastex6}
\pdfoutput=1 
\usepackage{amsmath,amsthm,amssymb,amsfonts}
\usepackage{float,xspace,makecell}
\usepackage[figuresright]{rotating}
\usepackage{placeins}

\newcommand\mosfit{{\tt MOSFiT}\xspace}
\newcommand\mist{{\tt MIST}\xspace}
\newcommand\flash{{\tt FLASH}\xspace}

\newcommand\emcee{{\tt emcee}\xspace}

\newcolumntype{C}[1]{>{\centering\let\newline\\\arraybackslash\hspace{0pt}}m{#1}}

\shorttitle{Weighing Black Holes using Tidal Disruption Events}
\shortauthors{Mockler et al.}

\begin{document}

\title{Weighing Black Holes using Tidal Disruption Events}

\author{Brenna Mockler\altaffilmark{1,2}, James Guillochon\altaffilmark{3}, Enrico Ramirez-Ruiz\altaffilmark{1,2}}
\affil{$^1$Department of Astronomy and Astrophysics, University of California, Santa Cruz, CA 95064, USA\\
$^2$DARK, Niels Bohr Institute, University of Copenhagen, Blegdamsvej 17, 2100 Copenhagen, Denmark\\
$^3$Harvard-Smithsonian Center for Astrophysics, 60 Garden St., Cambridge, MA 02138, USA}
\email{bmockler@ucsc.edu}

\begin{abstract}
While once rare, observations of stars being tidally disrupted by supermassive black holes are quickly becoming commonplace. To continue to learn from these events it is necessary to robustly and systematically compare our growing number of observations with theory. We present a tidal disruption module for the Modular Open Source Fitter for Transients (\mosfit) and the results from fitting 14 tidal disruption events (TDEs). Our model uses \flash simulations of TDEs to generate bolometric luminosities and passes these luminosities through viscosity and reprocessing transformation functions to create multi-wavelength light curves. It then uses an MCMC fitting routine to compare these theoretical light curves with observations. We find that none of the events show evidence for viscous delays exceeding a few days, supporting the theory that our current observing strategies in the optical/UV are missing a significant number of viscously delayed flares. We find that the events have black hole masses of $10^6 - 10^8 M_{\odot}$, and that the masses we predict are as reliable as those based on bulk galaxy properties. We also find that there is a preference for stars with mass $< 1 M_{\odot}$, as expected when low-mass stars greatly outnumber high-mass stars.
\end{abstract}


\section{Introduction}\label{sec:intro}
One of the most promising avenues for studying black holes in quiescent galaxies is through tidal disruption events (TDEs). Unlucky stars that pass too near a black hole are torn apart, lighting up previously dormant black holes \citep{Rees:1988a} and encoding the resultant light curves with a wealth of information about the  nature of disruptor and disruptee. The initial disruption tests how stars behave under the presence of strong gravity \citep{Kobayashi:2004a,Guillochon:2009a}.	The shape of the light curve includes clues about the mass and spin of the black hole \citep{Evans:1989a,Kesden:2012a,Cheng:2014a,Tejeda:2017a}, as well as the properties of the star \citep{Lodato:2009a,Haas:2012a,Law-Smith:2017a}, and the mechanics of the disruption and accretion processes \citep{Rosswog:2009a,Ayal:2000a,Bonnerot:2016a}. 

For a TDE to be observable, the tidal disruption radius, $R_{\rm t} \equiv (M_{\rm h}/M_{\ast})^{1/3} R_{\ast}$ of a star of mass $M_{\ast}$ and radius $R_{\ast}$ by a black hole of mass $M_{\rm h}$ must be outside the gravitational radius of the black hole \citep[e.g.][]{MacLeod:2012a}, else the black hole will swallow the star whole. For most stars, black holes $\lesssim~10^8 M_{\odot}$ are the most likely disruptors. This makes TDEs all the more exciting, as they are probing lower mass black holes that are otherwise difficult to study, and whose mass determinations are uncertain.

The fallback rate and the peak timescale of TDEs are dependent on the mass of the disrupting black hole, the mass of the star, and the stellar structure of the star \citep{Lodato:2009a,Guillochon:2013a}. Because the dependence on the mass and radius of the star largely cancel one another out on the main sequence, the peak timescale is sensitive to the mass of the black hole.  Thus, if a TDE's luminosity follows the fallback rate \citep[i.e. is ``prompt''][]{Guillochon:2015b}, the light curve can be used to measure the black hole's mass and the properties of the disrupted star. In order for the luminosity to follow the fallback rate, the stellar debris that initially returns on highly eccentric orbits must circularize on a timescale that is shorter than the fallback timescale \citep{Shiokawa:2015a,Bonnerot:2016a,Hayasaki:2016a}. As we show here, the optical and UV events that we modeled all require prompt circularization, suggesting that we can use their light curves to acquire reliable black hole mass measurements. 

New TDEs have been  uncovered  at a steady rate in recent years and the rate of discoveries will continue to increase. As such, it has become imperative to be able to systematically quantify the key variables responsible for shaping TDE light curves  so that we can compare these variables across events and develop a statistical understanding of the  physical ingredients at play. To facilitate this, it is important for TDE data to be accessible, and the \textit{Open TDE Catalog} \citep{Auchettl:2017a,Guillochon:2017a} is aiming to do this by collecting TDE data and hosting it online in a standardized format. To compare and contrast between different TDEs it is important to fit the events consistently, and to this end in this paper we introduce a theoretical model for fitting TDEs as part of \mosfit, the modular Open-Source Fitter for Transients \citep{Guillochon:2017a}. This model has been implemented in \mosfit and is available immediately.

Along with the model we present fits to the optical and UV data of 14 TDEs from the \textit{Open TDE Catalog}. Using \mosfit we are able to extract posterior distributions for key parameters, most notably the black hole mass. We attempt to capture the broad features of a TDE while minimizing the number of free parameters in our model. Our model ingredients are outlined in Section~\ref{sec:mosfit} and our TDE sample is described in Section~\ref{sec:fits}. Our black hole mass estimates are presented in Section~\ref{sec:bhs} along with a detailed comparison with those derived using other methods.  In Section~\ref{sec:disc} we discuss how the  posteriors from our fits can help inform TDE emission models and presents a summary of our findings.

\section{Method}\label{sec:method}
The tidal disruption model in \mosfit uses \flash simulations of the mass fallback rate \citep{Guillochon:2013a} as inputs to fit data of TDEs. It is modeled similarly to {\tt TDEFit}, a code for fitting tidal disruption events, originally described in \citet{Guillochon:2014a}, but excludes a few features of that code that will be ported to future versions of the \mosfit model (see Section~\ref{sec:summary}).
In the sections that follow we provide a detailed description of the model components along with a brief overview of the fitting procedure. 

\subsection{MOSFiT Modules}\label{sec:mosfit}

The \mosfit platform sub-divides the components of a model into independent modules such that common operations for fitting transients can be utilized by various transient types. This means any new model implemented in \mosfit re-uses many existing modules, reducing the chance of coding errors and improving overall performance. Below, we describe the new modules added to \mosfit specifically created for modeling TDEs, which include new {\it engine} (source of radiant emission), {\it transform} (reprocessing of radiant emission), and {\it photosphere} (conversion of bolometric flux to a distribution of flux as a function of wavelength) modules.

\begin{table}[h!]\label{table:parameter ranges}
    \renewcommand{\thefootnote}{\arabic{footnote}}
    \footnotesize
    \setlength\tabcolsep{3pt}
    \renewcommand{\arraystretch}{1.4}
\begin{tabular}{cccc} 
\hline
Parameter & Prior & Min & Max \\
\hline
$M_{\rm h} (M_{\odot})$  & Log & $10^5$ & $5 \times 10^8$ \\
$b$ (scaled impact parameter$^{\rm a}$) & Flat & 0 & 2 \\
$M_{\ast} (M_{\odot})$  & Kroupa & 0.01 & 100 \\
$\epsilon$ (efficiency) & Flat & 0.005 & 0.4 \\ 
$\rm $R$_{\rm ph0}$ (photosphere power law constant)  & Log & $10^{-4}$ & $10^{4}$ \\
$l$ (photosphere power law exponent) & Flat & 0 & 4 \\
$t_{\rm first~fallback}$ (days since first detection$^{\rm bc}$) & Flat & -500 & 0 \\
$T_{\rm viscous}$ (days) & Log & $10^{-3}$ & $10^5$ \\
\hline
\end{tabular}
\\
{
$^{\rm a}$The parameter $b$ is a proxy for $\beta$ as the relationship between $\beta$ and $\Delta M$ bound to the black hole differs for different $\gamma$. Minimum disruptions for both $\beta_{5/3}$ and $\beta_{4/3}$ correspond to $b$ = 0 and full disruptions for both $\beta$ correspond to $b$ = 1. Disruptions with $b$ = 2 correspond to $\beta_{5/3} = 2.5$ and $\beta_{4/3} = 4.0$ respectively.\\
$^{\rm b}$For our fit of iPTF16fnl we narrowed the range of $t_{\rm disruption}$ as \mosfit was having difficulty isolating the relatively short peak for that event, it is clear from the photometry that $t_{\rm first~fallback}$ is $\ll 500$ days before the first observation.\\
$^{\rm c}$ The parameter $t_{\rm first~fallback}$ is different from the time of disruption. For any combination of disruption parameters ($\beta$, $\gamma$) there exists a fixed time between $t_{\rm disruption}$ and $t_{\rm first~fallback}$. This delay can be affected by the precession of debris out of the original orbital plane, however it does not affect the determination of $M_h$ because the mass-energy distribution remains intact during this delay (see Section~\ref{sec:lum follows fallback}).
}
\caption{Here we list the parameters and priors used in our model. Where the listed prior is `Log', the natural logarithm was used.}
\end{table}

\subsubsection{Fallback Engine}
The {\it engine} for the TDE model comes from converting the fallback rate of material onto the black hole post-disruption directly to a bolometric flux via a constant efficiency parameter $\epsilon$. To model this process we used hydrodynamical simulations of polytropic stars tidally disrupted by supermassive black holes (SMBHs) \citep{Guillochon:2013a}. Polytropic stars are stars whose equation of state is defined by $P \propto \rho ^ \gamma$. The parameter $\gamma$ is the polytropic index -- colloquially the `polytrope.' Stars of different masses are better represented by different polytropes, we take stars with mass $\leq 0.3 M_{\odot}$ and mass $\geq 22 M_{\odot}$ to be represented by 5/3 polytropes ($\gamma = 5/3$) while stars with masses between $1 M_{\odot}$ and 15 $M_{\odot}$ are modeled as 4/3 polytropes ($\gamma = 4/3$). For stars in the transition ranges ($0.3 M_\odot$ -- $1 M_\odot$, $15 M_\odot$ -- $22 M_\odot$), we use hybrid fallback functions that smoothly blend between the 4/3 and 5/3 polytopes, the details of which are described later in this section. The simulations were run for a wide range of impact parameters ($ \beta = R_{\rm t}/R_{\rm p}$, $R_{\rm p}$ is the pericenter radius), varying from interactions that barely disrupted the star to interactions with $\beta$ values significantly larger than what is needed for full disruption. Stars are considered to be fully disrupted when no surviving core remains post-disruption, which for SMBH encounters yields a fallback mass $\Delta M = M_\ast/2$. Because both the mass of the black hole and the mass of the star enter into the rate of fallback as simple scaling parameters \citep{Guillochon:2013a, Guillochon:2015a}, all simulations were run with $M_{\rm h} = 10^6 M_{\odot}$ and $M_{\ast} = 1 M_{\odot}$. 

The hydrodynamical simulations provide us with the distribution of debris mass $dm/de$ as a function of specific binding energy $e$ after it is torn apart. This distribution is dependent on the structure of the star, a feature that is particularly important when fitting the shape of the light curve and its power-law decline at late times. To obtain the fallback rate $dm/dt = \dot{M}$, $dm/de$ is converted into a mass distribution in time using $de/dt$ calculated from keplerian orbital dynamics,

\begin{equation}
\label{eqn:period}
T = \frac{\pi \mu}{\sqrt{2}} (- e)^{-3/2}
\end{equation}

\begin{equation}
\label{eqn:dedt}
de/dt = \frac{(2 \pi \mu )^{2/3}}{3 t^{5/3}}.
\end{equation}

In the above equations, $e$ is the specific orbital energy, $\mu$ is the standard gravitational parameter, and $T$ is the orbital period of the bound debris that falls back onto the black hole.

Our model assumes that stars meet black holes on approximately zero-energy (parabolic) orbits, as is true for most tidal disruptions in galactic nuclei. This means that the energy of the bound stellar debris is only dependent on the potential of the SMBH. Using this simplification and taking a Taylor expansion of the potential of the SMBH at the surface of the star at the pericenter of the orbit, one finds

\begin{equation}
\label{eqn:specificenergy}
e \propto \frac{G M_{\rm h} R_{\ast}}{R_t^2} \propto M_{\rm h}^{1/3} M_{\ast}^{2/3} R_{\ast}^{-1}.
\end{equation}

The mass-energy distribution is related to the black hole mass and specific binding energy through $dm/de \propto M_{\rm h}/ 2e$ when the ratio of the black hole's mass to the star's mass is large \citep{Rees:1988a,Phinney:1989a}. Substituting Equation~\ref{eqn:specificenergy} into Equation~\ref{eqn:period} gives us the dependence between the timescale and the properties of the star and the black hole;  $t \propto M_h^{1/2} M_{\ast}^{-1} R_{\ast}^{3/2}$. Using this relation together with Equation~\ref{eqn:dedt}, we find that $de/dt \propto M_h^{-1/6} M_{\ast}^{5/3} R_{\ast}^{-5/2}$ and therefore $dm/dt = dm/de \times de/dt \propto M_h^{-1/2} M_{\ast}^{2} R_{\ast}^{-3/2}$. To summarize, the following relations relate the parameters of the star and black hole to the mass fallback rate, 

\begin{equation}
\label{eqn:lpeak}
\dot{M} \propto M_h^{-1/2} M_{\ast}^{2} R_{\ast}^{-3/2},
\end{equation}
\begin{equation}
\label{eqn:tpeak}
t(\dot{M}) \propto M_h^{1/2} M_{\ast}^{-1} R_{\ast}^{3/2}.
\end{equation}
Here we use $\dot{M}$ to denote the fallback rate, so $t(\dot{M})$ is the time of a given fallback rate. We will continue to use this notation throughout the rest of the paper.

After collecting $\dot{M}$ for various values of $\beta$ and $\gamma$, values for $\beta$, $M_{\ast}$ and $M_{\rm h}$ are input into the {\it fallback} module, which linearly interpolates in $\beta$-$M_\ast$ space (using the mapping between $M_\ast$ and $\gamma$ described above) to obtain fallback curves as a functions of both parameters. In order to provide accurate description for the light curve with $M_{\ast}$ and $M_{\rm h}$, we make use of the following scalings given in Equation~\ref{eqn:lpeak} and Equation~\ref{eqn:tpeak}. 

We use \citet{Tout:1996a} to get $R_{\ast}$ from $M_{\ast}$ for $M_{\ast} \geq 0.1 M_{\odot}$. Below that mass we assume that the radius is constant and use $R_{\ast, \rm Tout}(M_{\ast} = 0.1 M_{\odot}) \approx 0.1 R_{\odot}$, roughly the radius of Jupiter.

We also assume the stars are zero-age main-sequence stars (ZAMS) and that they have solar metallicity. Both the ZAMS and composition assumption as well as the assumption that the stars are represented by blends of 4/3 and 5/3 polytropes are simplifying assumptions that allow us to build this minimal model without introducing excessive numbers of free parameters. In future work we plan to use simulations of realistic stars for a wide range of ages and compositions as inputs into our fallback module \citep{Law-Smith:2017a}. 

At the end of the fallback module, we convert $\dot{M}$ to luminosity by assuming a constant efficiency $\epsilon$, which we allow to vary as a free parameter in our fitting procedure, yielding $L = \epsilon \dot{M} c^2$. This freedom allows us to remain agnostic about the physical mechanism driving this conversion, which can be sub-percent if originating from a stream-stream collision \citep{Jiang:2016a} or up to 42\% if the conversion occurs at the ISCO of a maximally-spinning black hole \citep{Beloborodov:1999a}. We also introduce a soft cut at the Eddington limit $L_{\rm Edd} \equiv 4 \pi G M_{\rm h} c / \kappa$ to prevent the radiated luminosity from exceeding this value (here $\kappa$ is the mean opacity to Thomson scattering assuming solar metallicity). This is motivated by both the fact that the peak bolometric luminosities derived observationally for optical/UV TDEs appear to be sub-Eddington \citep{Hung:2017a,Wevers:2017a} and that other accreting black hole systems (such as AGN) rarely show evidence for large thermal Eddington luminosity excesses.

\subsubsection{Viscous Delay}
The assumption that the luminosity closely follows the fallback rate is a bold assertion that, if correct, gives us a deterministic way to relate how stellar debris circularizes and how it accretes onto the black hole. We define a `viscous time' in this work, which encompasses the effects of time delays due to the circularization process as well as delays due to accretion through the disk surrounding the black hole.  If the viscous time about the black hole were short as compared to the fallback time, the accretion rate onto the black hole from the forming disk $\dot{M}_{\rm d}$ should be equal to the fallback rate $\dot{M_{\rm fb}}$. It is likely that once debris reaches the disk the time it will take to accrete onto the black hole will be much shorter than the fallback time. The orbital timescale at the edge of the disk ($\sim 2 R_{\rm p}$) is much smaller than the original orbital timescales of the debris (Equation~(\ref{eqn:period})), and therefore viscous processes in the disk have many (disk) orbital timescales over which to move debris inward. However, as has been found in several numerical works \citep{Guillochon:2014a,Shiokawa:2015a,Bonnerot:2016a,Hayasaki:2016a}, circularization about the black hole might be very ineffective, resulting in viscous times that are potentially hundreds of times longer than the orbital period of the most-bound debris \citep{Cannizzo:1990a,Guillochon:2015b,Dai:2015a}. This would result in a central accretion disk with $R \approx 2 R_{\rm p}$ that is starved of mass, with much of the mass being held aloft for long periods of time in an elliptical superstructure \citep{Ramirez-Ruiz:2009a,Guillochon:2014a}. While the exact details of how matter is received by the disk and then later accreted by the black hole remain elusive \citep{Sadowski:2016a}, the primary effect of the viscous slow-down is likely well-approximated as a ``low-pass" filter on the fallback rate,
\begin{equation}
\dot{M}_{\rm d}(t) = \dot{M}_{\rm fb}(t) - M_{\rm d}(t)/T_{\rm viscous},
\end{equation}
where the elliptical disk that forms acts as a reservoir where a mass $M_{\rm d}$ remains suspended outside of the black hole's horizon for roughly a viscous time. The solution to this expression is
\begin{equation}
\dot{M}_{\rm d}(t) = \frac{1}{T_{\rm viscous}}\Big(e^{-t/T_{\rm viscous}}\int^t_0 e^{t'/T_{\rm viscous}}\dot{M}_{\rm fb}(t')dt'\Big),
\end{equation}
which shows that the accretion rate exponentially approaches the fallback rate after a viscous time. We implement the above expression in our viscous module, inputting the luminosities from our fallback module through the transform, which yields viscously-delayed luminosities that are used to compute light curves.

\subsubsection{Photosphere}
Regardless of the process or combination of processes responsible for generating the emission, the kinetic energy of the returning debris must eventually be dissipated in order to be observed. Even if some energy is deposited by circularization at large distances \citep{Piran:2015b}, the energy will be primarily dissipated by processes that operate closest to the black hole simply because the velocities there are the greatest. However, this would imply most of the radiation would be emitted at very high energies (X-rays), and instead we observe many TDEs with significant (and sometimes dominant) optical/UV flux. A reprocessing layer, either static or outflowing \citep{Miller:2015a,Metzger:2016b}, can help explain the observed emission by reprocessing the luminosity generated by the various dissipation processes at play \citep{Loeb:1997a,Ulmer:1998a,Bogdanovic:2004a,Guillochon:2014a,Jiang:2016a,Coughlin:2014a,Strubbe:2009a}. The reprocessing of the radiation has also been used to successfully explain the line ratios observed in PS1-10jh \citep{Roth:2016a,Gaskell:2014a}. In this work we assume a simple blackbody photosphere for the reprocessing layer, so that the observed flux becomes
\begin{equation}
F_\nu=\frac{2\pi h\nu^3}{c^2}   \frac{1}{\exp(h\nu/kT_{\rm eff})-1}    \frac{R_{\rm phot}^2}{D^2},
\end{equation}
with an effective blackbody temperature
\begin{equation}
T_{\rm eff}=\bigg{(} \frac{L}{4\pi \sigma_{\rm SB}R_{\rm phot}^2} \bigg{)}^{1/4}.
\end{equation}
In the above equations, $F_\nu$ is the specific flux, $R_{\rm phot}$ is the photospheric radius, $D$ is the distance from the source, $L$ is the bolometric luminosity from our fit, and $T_{\rm eff}$ is the temperature of the photosphere.
Most observations of TDEs have thermal temperatures that don't exhibit significant variation. For blackbody emission, the radius must increase as the luminosity (and $\dot{M_{\rm fb}}$) increase, and decrease as the luminosity decreases, in order for the temperature not to change significantly as the luminosity evolves. This simple behavior also explains the rise in temperatures at late times as the photospheric radius decreases and the bulk of the observed radiation shifts to higher energies. To model this dependence we assume that the radius of the photosphere has a power law dependence on the luminosity and fit for both the power law exponent $l$ and radius normalization $R_{\rm ph0}$,
\begin{equation}
\label{eqn:R to L}
R_{\rm phot} = R_{\rm ph0} a_p (L/L_{\rm edd})^l.
\end{equation}
Here $a_{\rm p}$ is the semi-major axis of the accreting mass at peak $\dot{M}_{\rm fb}$. This provides a reasonable typical scaling for the radius of the photosphere, with a minimum photosphere size set by $R_{\rm isco}$ and a maximum photosphere size set by the semi-major axis of the accreting mass.

One of the appealing aspects of this photosphere model is that it remains agnostic towards the mechanism ultimately responsible for generating the luminosity, but does make a number of simplifying assumptions regarding the source function of the radiation. In particular, it assumes that all of the radiation is efficiently thermalized at the scale of the photosphere radius. The resultant spectrum is compatible with what one would expect from a ``veiled'' TDE \citep{Auchettl:2017a}, and, as such,  this model cannot reproduce the x-ray emission that is observed in a small fraction of TDEs found in optical surveys \citep[e.g. ASASSN-14li,][]{Miller:2015b}. In the future, we plan to include an accretion disk module which will be used to describe the x-ray emission that sometimes is observed to accompany optical/UV TDEs \citep{Auchettl:2017a}.

\section{Light Curve Fits}\label{sec:fits}
The characteristics of the population of TDEs as a whole can be derived by fitting a significant fraction of the existing TDE candidates to a shared model. In what follows we describe the data used in this study as well as the results from the fitting procedure.   

\begin{figure*}[h!]
\centering
\includegraphics[scale=0.39]{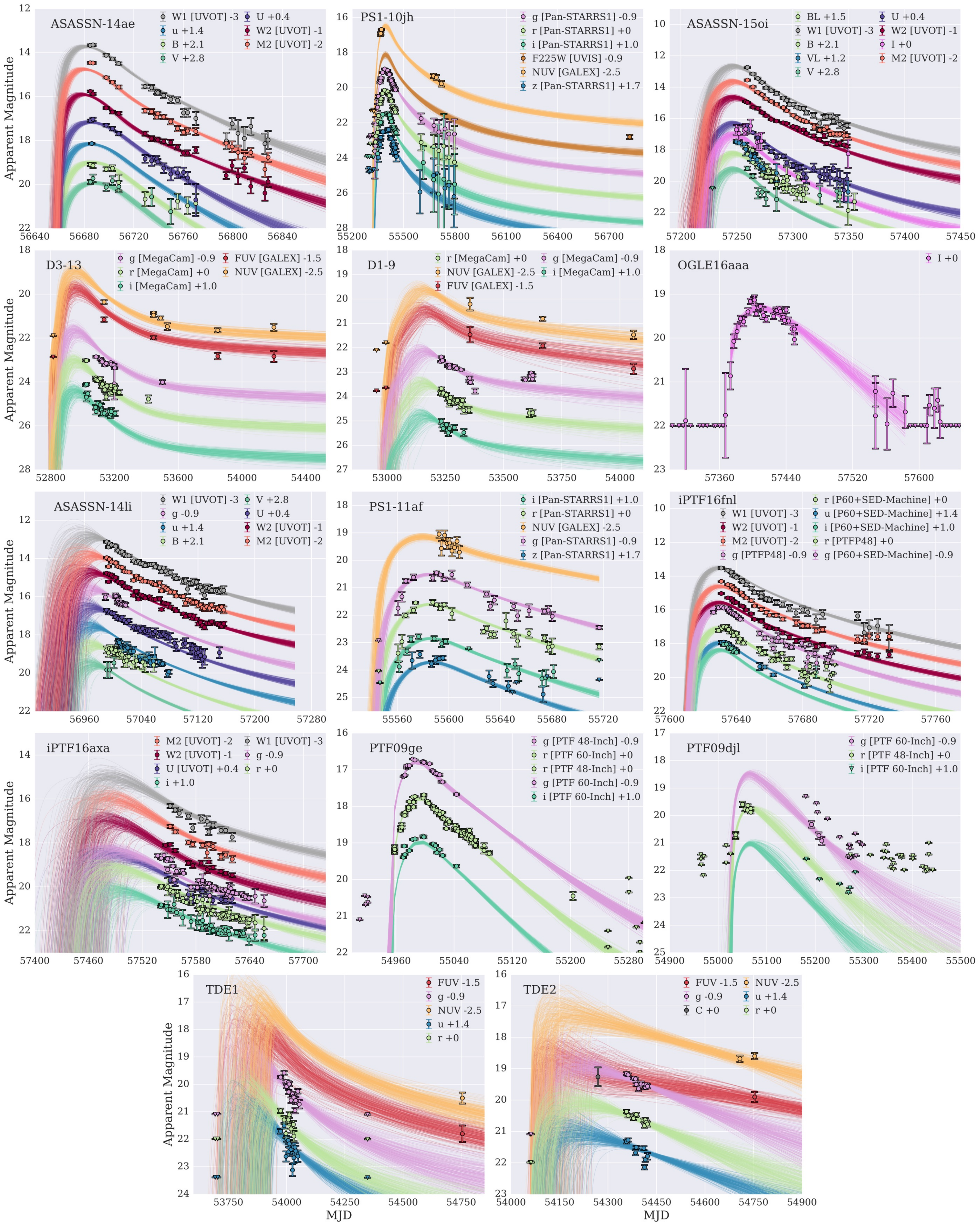} 
\caption{Ensembles of TDE light curves each constructed from the posterior parameter distribution. The multicolor detections and associated upper limits are plotted for all selected TDEs.}
\label{fig:lc grid}
\end{figure*}

\subsection{Data Selection}
The data from our fits is public and can be found on the Open TDE Catalog\footnote{\url{https://tde.space}}. There does not exist a single agreed upon test for classifying a transient as a TDE, and therefore multiple clues must be taken together to determine the likelihood that a transient is in fact the result of a TDE. First of all, astrometry must place the transient near the center of its host galaxy. Next, unique light curve features (blue optical/UV colors, minimal color evolution, and a large brightening above the quiescent level) are used to separate TDEs from other transients occurring in the cores of galaxies such as AGN flares \citep[e.g.,][]{Gezari:2009a}. Spectra of the events, in particular transient broad features of hydrogen and helium \citep{Arcavi:2014a}, are also used to separate the events from other phenomena, particularly supernovae. Finally, we theoretically expect the bolometric light curves to have a power law decline at late times \citep{Rees:1988a,Lodato:2012a,Guillochon:2013a}, as opposed to an exponential decline that might be better associated with nuclear decay and thus a supernova origin.

In selecting data we were limited by the confines of our current model. For example, we currently do not fit x-ray radiation, and therefore we required events in our sample to have bolometric luminosities dominated by emission in the optical/UV. In addition to this, our current model can only reproduce light curves with a single temporal component, and we are thus unable to fit events such as ASASSN-15lh that have a significant late time re-brightening that might arise from an emerging accretion disk \citep{Margutti:2017a}. From this subset of TDEs we first chose events which had either observations of the light curve peak or near-peak early time upper limits. The peak timescale of most TDEs is $\lesssim$ 1 year, therefore we defined near-peak upper limits to be within 1 year of the first observed data point.
All of the TDEs in our sample except iPTF16axa, TDE1 and TDE2 fall into this category. We additionally included events with detailed observations of the decline ($\geq 3$ data points in each band over the course of a peak timescale) in at least three optical/UV bands, even if these events did not have peak observations or near-peak upper limits (such as iPTF16axa, TDE1 and TDE2). With sufficient color information, \mosfit is able to constrain the bolometric luminosity curve and therefore also the peak timescale.

Here we include a list of the events we fit: PS1-10jh \citep{Gezari:2012a,Gezari:2015a}, PS1-11af \citep{Chornock:2014a}, PTF09djl \citep{Arcavi:2014a}, PTF09ge \citep{Arcavi:2014a}, iPTF16fnl \citep{Blagorodnova:2017a,Brown:2017a}, iPTF16axa \citep{Hung:2017a}, ASASSN-14li \citep{Holoien:2016a, Brown:2016b}, ASASSN-15oi \citep{Holoien:2016b}, ASASSN-14ae \citep{Holoien:2014a, Brown:2016c}, OGLE16aaa \citep{Wyrzykowski:2017a}, D1-9 \citep{Gezari:2008a}, D3-13 \citep{Gezari:2008a}, TDE1 \citep{van-Velzen:2011a}, and TDE2 \citep{van-Velzen:2011a}.

\subsection{Fitting Procedure}
\mosfit currently uses a variant of the \emcee ensemble-based MCMC routine \citep{Foreman-Mackey:2013a} to find the combinations of parameters that yield the highest likelihood matches for a given input model \citep{Guillochon:2017d}, where model errors are fitted simultaneously with model parameters by the variance parameter $\sigma$. To quantify how well the various combinations of parameters in the model fit each light curve, \mosfit uses the Watanabe-Akaike information criteria \citep{Watanabe:2010a} (WAIC), also known as the widely applicable Bayesian criteria. This is used in place of the total evidence of the model: for objective functions where the likelihood function is not analytic and separable (such as in this semi-analytic model), it is difficult to evaluate the evidence exactly. While the WAIC score does not directly scale with the evidence, it is correlated with it, and can be used to rank fits between models \citep[see Section~7 of][]{Gelman:2014a}. The WAIC is evaluated as follows,
\begin{equation}
\mathrm{WAIC} = \overline{\textrm{log } p_n} - \mathrm{var}(\textrm{log } p_n),
\end{equation}
where $\overline{p_n}$ is the mean log likelihood score and $\mathrm{var}(\textrm{log } p_n)$
its variance.

\begin{table*}[ht]
\centering
    \renewcommand{\thefootnote}{\arabic{footnote}}
    \footnotesize
    \setlength\tabcolsep{2.1pt}
    \renewcommand{\arraystretch}{1.5}
\begin{tabular}{cccccccccccccc}
\hline
TDE & $M_{\rm h}$ & $\beta$ & $M_{\ast}$ & $\gamma$ &$\epsilon$ & $\rm log_{\rm 10}($R$_{\rm ph0})$ & $l$ & $t_{\rm peak}$ & $T_{\rm viscous}$ & $\sigma$ & WAIC & PSRF & Source \\
& ($10^6 M_{\odot}$) & & ($M_{\odot}$) & & & & & (days) & (days) & & & &\\
\hline
PS1-10jh & $17_{-1}^{+2}$ & $0.899_{-0.005}^{+0.006}$ & $0.101_{-0.002}^{+0.002}$ & 5/3 & $0.09_{-0.02}^{+0.03}$ & $0.8_{-0.1}^{+0.1}$ & $1.4_{-0.1}^{+0.1}$ & $110_{-2}^{+7}$ & $0.08_{-0.08}^{+0.85}$ & $0.11_{-0.01}^{+0.01}$ & 200 & 1.10 & 1, 2 \\
PS1-11af & $3.7_{-0.4}^{+0.5}$ & $0.90_{-0.01}^{+0.03}$ & $0.101_{-0.003}^{+0.009}$ & 5/3 & $0.020_{-0.003}^{+0.005}$ & $1.4_{-0.1}^{+0.1}$ & $1.4_{-0.1}^{+0.1}$ & $54_{-3}^{+3}$ & $0.2_{-0.2}^{+1.4}$ & $0.06_{-0.01}^{+0.01}$ & 231 & 1.07 & 3 \\
PTF09djl & $2.6_{-0.5}^{+1.0}$ & $0.86_{-0.09}^{+0.06}$ & $0.11_{-0.02}^{+0.13}$ & 5/3 & $0.1_{-0.1}^{+0.1}$ & $1.1_{-0.4}^{+0.8}$ & $2.1_{-0.4}^{+0.8}$ & $54_{-7}^{+10}$ & $0.2_{-0.2}^{+2.9}$ & $0.01_{-0.01}^{+0.02}$ & 120 & 1.25 & 4 \\
PTF09ge & $3.6_{-0.5}^{+0.8}$ & $1.1_{-0.3}^{+0.1}$ & $0.10_{-0.01}^{+0.07}$ & 5/3 & $0.008_{-0.002}^{+0.003}$ & $3.5_{-0.2}^{+0.2}$ & $1.97_{-0.09}^{+0.10}$ & $59_{-4}^{+14}$ & $0.04_{-0.04}^{+0.58}$ & $0.09_{-0.01}^{+0.01}$ & 94 & 1.81 & 4\\
iPTF16fnl & $1.7_{-0.2}^{+0.2}$ & $0.91_{-0.02}^{+0.05}$ & $0.101_{-0.004}^{+0.008}$ & 5/3 & $0.007_{-0.002}^{+0.002}$ & $1.0_{-0.1}^{+0.1}$ & $1.7_{-0.1}^{+0.1}$ & $37_{-2}^{+2}$ & $0.04_{-0.04}^{+0.32}$ & $0.21_{-0.02}^{+0.01}$ & 142 & 1.12 & 5, 6 \\
iPTF16axa & $2.5_{-0.9}^{+1.3}$ & $0.94_{-0.07}^{+0.08}$ & $1.0_{-0.2}^{+0.8}$ & 4/3 & $0.02_{-0.01}^{+0.03}$ & $0.8_{-0.2}^{+0.2}$ & $0.6_{-0.1}^{+0.1}$ & $62_{-9}^{+13}$ & $0.2_{-0.2}^{+3.4}$ & $0.21_{-0.01}^{+0.02}$ & 139 & 1.32 & 7\\
ASASSN-14li & $9_{-3}^{+2}$ & $0.90_{-0.04}^{+0.15}$ & $0.2_{-0.1}^{+0.1}$ & 5/3 & $0.2_{-0.1}^{+0.1}$ & $-0.2_{-0.1}^{+0.2}$ & $1.9_{-0.3}^{+0.4}$ & $95_{-15}^{+14}$ & $0.1_{-0.1}^{+2.1}$ & $0.14_{-0.01}^{+0.01}$ & 283 & 1.11 & 8\\
ASASSN-15oi & $4_{-1}^{+1}$ & $0.91_{-0.02}^{+0.06}$ & $0.11_{-0.01}^{+0.04}$ & 5/3 & $0.018_{-0.005}^{+0.010}$ & $1.3_{-0.2}^{+0.3}$ & $2.1_{-0.2}^{+0.1}$ & $60_{-7}^{+8}$ & $0.04_{-0.04}^{+0.53}$ & $0.20_{-0.02}^{+0.02}$ & 73 & 1.10 & 9 \\
ASASSN-14ae & $1.3_{-0.1}^{+0.1}$ & $1.04_{-0.06}^{+0.03}$ & $1.00_{-0.02}^{+0.02}$ & 4/3 & $0.006_{-0.001}^{+0.003}$ & $1.8_{-0.1}^{+0.1}$ & $1.3_{-0.1}^{+0.1}$ & $37_{-2}^{+3}$ & $0.05_{-0.05}^{+0.52}$ & $0.12_{-0.01}^{+0.02}$ & 97 & 1.06 & 10\\
OGLE16aaa & $3.0_{-0.8}^{+1.2}$ & $0.81_{-0.09}^{+0.10}$ & $0.2_{-0.1}^{+0.2}$ & 5/3 & $0.2_{-0.1}^{+0.2}$ & $1.3_{-0.6}^{+0.7}$ & $1.7_{-0.4}^{+0.6}$ & $67_{-13}^{+13}$ & $0.1_{-0.1}^{+2.3}$ & $0.11_{-0.02}^{+0.03}$ & 38 & 1.16 & 11\\
D1-9 & $66_{-10}^{+7}$ & $1.2_{-0.2}^{+0.2}$ & $7_{-3}^{+5}$ & 4/3 & $0.11_{-0.07}^{+0.16}$ & $-0.5_{-0.8}^{+1.0}$ & $3.4_{-0.8}^{+0.4}$ & $212_{-33}^{+42}$ & $0.1_{-0.1}^{+2.4}$ & $0.24_{-0.06}^{+0.11}$ & 96 & 1.44 & 12\\
D3-13 & $30_{-3}^{+3}$ & $1.8_{-0.8}^{+0.1}$ & $7_{-4}^{+17}$ & 4/3 & $0.2_{-0.1}^{+0.1}$ & $-1.2_{-0.1}^{+0.1}$ & $3.8_{-0.4}^{+0.2}$ & $131_{-11}^{+71}$ & $0.1_{-0.1}^{+3.3}$ & $0.27_{-0.03}^{+0.03}$ & 37 & 1.10 & 12\\
TDE1 & $3_{-1}^{+3}$ & $0.84_{-0.08}^{+0.13}$ & $0.1_{-0.1}^{+0.3}$ & 5/3 & $0.2_{-0.1}^{+0.2}$ & $0.2_{-0.3}^{+0.3}$ & $0.7_{-0.2}^{+0.3}$ & $76_{-25}^{+45}$ & $0.2_{-0.2}^{+5.8}$ & $0.20_{-0.03}^{+0.03}$ & 38 & 1.16 & 13 \\
TDE2 & $1.9_{-0.6}^{+1.4}$ & $1.0_{-0.1}^{+0.3}$ & $0.3_{-0.1}^{+0.3}$ & 5/3 & $0.2_{-0.1}^{+0.1}$ & $0.8_{-0.2}^{+0.1}$ & $1.5_{-0.5}^{+0.9}$ & $50_{-12}^{+36}$ & $0.2_{-0.2}^{+9.4}$ & $0.12_{-0.02}^{+0.02}$ & 53 & 1.19 & 13 \\
\hline
\vspace{-1.3em}
\end{tabular}
{
$^1$\citet{Gezari:2012a}, $^2$\citet{Gezari:2015a}, $^3$\citet{Chornock:2014a}, $^4$\citet{Arcavi:2014a}, $^5$\citet{Blagorodnova:2017a}, 
$^6$\citet{Brown:2017a}, $^7$\citet{Hung:2017a}, $^8$\citet{Holoien:2016a}, $^9$\citet{Holoien:2016b}, $^{10}$\citet{Holoien:2014a},
$^{11}$\citet{Wyrzykowski:2017a},$^{12}$\citet{Gezari:2008a}, $^{13}$\citet{van-Velzen:2011a}
}

\caption{ Here we list best fit parameters for all light curves with $1\sigma$ error bars.}
\label{table:3sigma parameters}
\end{table*}

In addition to measuring the goodness of fit, it is important to ascertain whether or not a fit has converged. To this end, we use the Gelman-Rubin metric, or Potential Scale Reduction Factor (PSRF, signified with $\hat{R}$) to gauge convergence \citep{Gelman:1992a}. This metric measures how well mixed each individual chain is as well as the degree of mixture between the different chains \citep[for the definition, see][]{Guillochon:2017d}.

For this multi-parameter model we used the maximum of the PSRFs computed for each parameter, so that the convergence of each fit was determined by the parameter with the slowest convergence. We attempted to run all of our fits until they reached a PSRF $\leq 1.2$ (ensuring that the walkers are well-mixed within the regions of convergence \citep{Brooks:1998a}, however this was not possible for every fit. The 4 events with PSRF $>1.2$ were refit multiple times, and continued to converge to the solutions we present here. For the work presented in this paper a minimum of 200 walkers and 30,000 iterations were used to recover the distribution of model fits.

\subsection{Results}\label{sec:results}
\begin{figure*}[t]
\vspace{-2mm}
\centering
\includegraphics[scale=0.255]{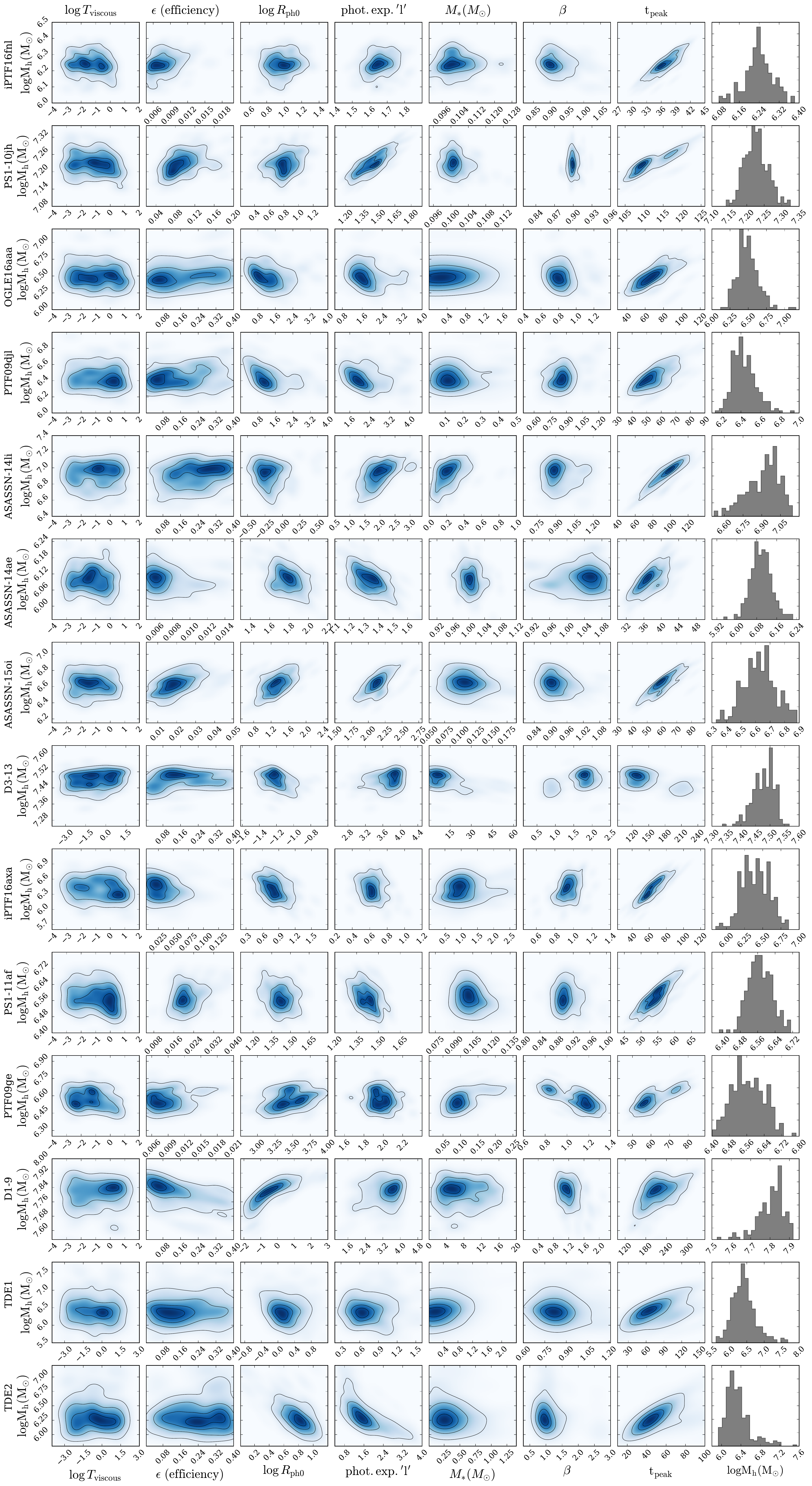} 
\caption{Posterior distributions of model parameters in the fit for each event as a function of $M_{\rm h}$. All logarithms are base 10. We include 0.5, 1, 1.5 and 2$\sigma$ contours for the 2-dimensional distributions -- these correspond to where 0.1175, 0.393, 0.675, and 0.865 of the 2D volume is contained. The plot shows that, for most events, $t_{\rm peak}$ (not itself a model parameter) correlates strongly with $M_{\rm h}$.}
\label{fig: 2D hist Mh vs params}
\end{figure*}

We show the results of the light curve fits in Figure~\ref{fig:lc grid}, the best fit parameter values in Table~\ref{table:3sigma parameters}, and the posterior distributions of the walkers in Figure~\ref{fig: 2D hist Mh vs params}. In Figure~\ref{fig:lc grid}, the ensemble of light curves from the final walker positions are plotted. Although the model priors allow for long viscous times, the light curves of highest likelihood continue to closely follow the fallback rates. The viscous timescales and $t_{\rm peak}$ values are shown in Table~\ref{table:3sigma parameters}. The preferred viscous delays are less than 1\% of $t_{\rm peak}$ for all events modeled in this work; this preference is visible in the first column of panel plots in Figure~\ref{fig: 2D hist Mh vs params}. The minimal viscous delay of these events allows us to obtain precise black hole mass measurements as the luminosity evolution is still best described using the fallback rate, and the primary dependence of $t_{\rm peak}$ is upon $M_{\rm h}$ (as shown in Equation~(\ref{eqn:tpeak})). 

\begin{table}[h!]
\centering 
    \renewcommand{\thefootnote}{\arabic{footnote}}
    \footnotesize
    \setlength\tabcolsep{2.5pt}
    \renewcommand{\arraystretch}{1.4}
\begin{tabular}{cccccccccccccc}
\hline
parameter & error & scale \\ 
\hline
$M_{\rm h}$ & $\pm 0.2$ & log$_{\rm 10}$ \\
$\beta$ & $\pm 0.35$ & linear \\
$M_{\ast}$ & $\pm 0.66$ & log$_{\rm 10}$\\
$\epsilon$ & $\pm 0.68$ & log$_{\rm 10}$\\
$R_{\rm ph0}$ & $\pm 0.4$ & log$_{\rm 10}$\\
$l$ & $\pm 0.2$ & linear \\
$t_{\rm peak}$ & $\pm 15$ & linear\\
$T_{\rm viscous}$ & $\pm 0.10$ & log$_{\rm 10}$\\
\hline
\vspace{-1.3em}
\end{tabular}
\caption{We present estimates of the systematic error in each parameter. These estimates were obtained by running fits with an additional variable parameterizing the uncertainty in the mass-radius relation of the disrupted stars, and comparing the results to our original measurements. This mass-radius relation is likely our largest source of systematic error in measuring the mass of the black hole.}
\label{table:systematic error}
\end{table}

In the absence of good photometry around peak, early time upper limits can help constrain the peak timescale and therefore the corresponding black hole mass, as shown in the plots for events D1-9, D3-13, PTF09djl, ASASSN-14li, ASASSN-15oi and ASASSN-14ae. For events without early time information or near-peak upper limits we can still fit the data if there is detailed observations (in multiple bands) of the decline. The mass fallback rate and bolometric luminosity do not decline with a constant power law, and this helps \mosfit find fits to events with well-sampled photometry but without early time data. Good band coverage allows \mosfit to accurately pin down different sections of the SED and more accurately measure the bolometric luminosity. It also makes it possible to constrain the photosphere parameters $R_{ph0}$ and $l$ (the power law constant and exponent, as defined in Equation~(\ref{eqn:R to L})). 

For example, our initial fit to ASASSN-15oi was completed before we realized there existed an early time upper limit, however the black hole mass we measure with the addition of that upper limit is the same as what we found without it. The other parameters similarly maintained their previously measured values, the upper limit simply reduced the uncertainty in the measurements.

The light curves of the majority of the events in this sample have one clear peak and monotonically decrease afterwards, as is predicted by our current single-component model. These include PS1-10jh, PS1-11af, PTF09ge, PTF09djl, ASASSN-14ae, OGLE16aaa, D3-13, D1-9, iPTF16axa, and iPTF16fnl. These events are also seen to radiate most of their bolometric luminosity at UV/optical wavelengths. They resemble veiled TDEs, in which the accretion disk is likely obscured by an optically thick photosphere or wind \citep{Auchettl:2017a}. However, there are a few TDEs in this sample (ASASSN-14li, ASASSN-15oi) whose x-ray emission and late time light curves are not as well described by our current single-component model and likely require a secondary component to explain their late-time behavior.

As shown in Figure~\ref{fig: Lbol, Tphot, Rphot}, the radius of the reprocessing layer in our model decreases at late times. Once the photosphere has receded to the size of the accretion disk, we expect higher energy photons to start contributing and ultimately dominating the light curve. As the luminosity decreases, the radiation from the accretion disk is expected to soften, potentially shifting the peak of the emission back into the UV/optical bands. At the same time, as the photosphere recedes, less x-rays from the accretion disk are expected to be reprocessed, allowing us to observe them. These additional late-time components can change the decline of the light curve. Of this sample, it is possible that for ASASSN-14li \citep{Brown:2017a}, and ASASSN-15oi \citep{Gezari:2017} these additional components play a role in their late time light curves. 

Although we did not model the origin of x-ray emission in this work, ASASSN-14li shows significant energy emitted at these wavelengths, which could be explained by the presence of a partially obscured accretion disk. In addition to this, the late time data shows that the decline of the UV light curve slows and the UV luminosity remains fairly constant from $\sim350$ days after discovery to the final observations at $\sim600$ days after discovery \citep{Brown:2017a} (the late time host-subtracted data was not available at the time of this study and therefore we did not fit it). Similarly, new late time observations of ASASSN-15oi from $\sim250$ days after discovery show flat optical/UV luminosities. ASASSN-15oi also exhibits an increasing x-ray component during the same time period \citep{Gezari:2017,Holoien:2018a}. When we attempted to fit ASASSN-15oi's late time optical and UV data with our model we found the quality of the fit worsened significantly, with the WAIC score dropping from 73 to 17. Therefore the fit we present here does not include the late time component of the light curve. Another potential example of a two-component TDE in the literature is ASASSN-15lh \citep{Nicholls:2015,Leloudas:2016a}. If ASASSN-15lh is indeed a TDE, then it requires a secondary late time component to explain the behavior of its light curve.

\section{Black hole mass predictions}\label{sec:bhs}
As discussed in the previous section, events with well-observed peaks and data in multiple bands have well-constrained black hole masses. The distributions of black hole masses for each event are shown in the last column of Figure~\ref{fig: 2D hist Mh vs params}, and the 68\% confidence intervals are listed in Table~\ref{table:3sigma parameters}. Figure~\ref{fig: 2D hist Mh vs params} shows 2D histograms of all parameters plotted against black hole mass in order to see correlations between the different variables. The most obvious and consistent correlation is between the black hole mass and the time of peak. Nevertheless, we might expect other parameters to be mildly correlated with  black hole mass as well. For example, the efficiency ($\epsilon$), $\beta$, and the star mass all enter into the peak luminosity scaling relation with $M_{\rm h}$. However, when looking at columns 2, 5 and 6 in Figure~\ref{fig: 2D hist Mh vs params}, we see that none of these variables have a clear correlation with black hole mass--perhaps their combined influence dilutes their individual correlations with $M_{\rm h}$. 

\begin{figure}[h!]
\centering
\includegraphics[width=1 \linewidth,clip=true]{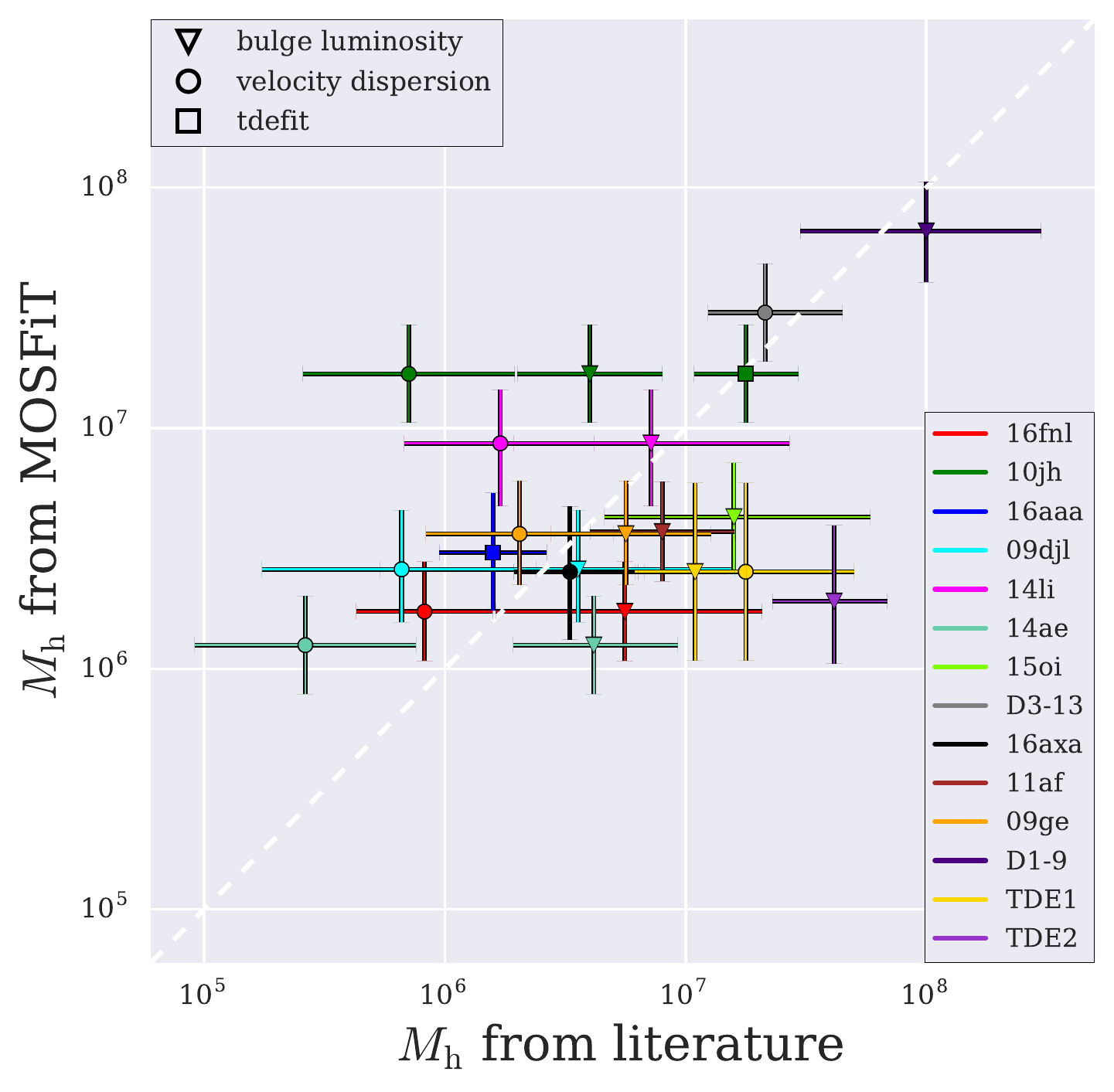} 
\caption{Comparison between the black hole mass estimates we derive from our model fits and those derived using the bulk properties of the host galaxy. The $M_h$ measurements from galactic properties come from the following sources: \citet{Arcavi:2014a, Blagorodnova:2017a, Brown:2017a, Chornock:2014a, Gezari:2008a, Guillochon:2014a, Holoien:2014a, Holoien:2016a, Holoien:2016b, Hung:2017a, Mendel:2014a, van-Velzen:2011a, Wevers:2017a, Wyrzykowski:2017a}. Measurements are averaged and errors are added in quadrature where multiple measurements using the same method exist for a single black hole. \mosfit error measurements include systematic error, literature error measurements include the intrinsic scatter in the relevant relation.}
\label{fig:Mh_vs_lit}
\end{figure}

The masses of the black holes we fit are all inferred by other mass estimation methods to be between $10^{5}$ and $10^{8}$ solar masses. In Figure~\ref{fig:Mh_vs_lit} we compare our results to mass measurements of the central black holes in the corresponding host galaxies using standard methods, and we find consistent results within reasonable errors (see Figure~\ref{fig:Msigma} for additional comparison with literature values). In this mass range, both the $M_{\rm h}-\sigma$ and $M_{\rm h}-L_{\rm bulge}$ relations suffer from significant uncertainty (see Section~\ref{sec:bhestimation}), therefore it is not surprising that masses derived using different scaling relations do not always agree within quoted errors. This makes the construction of an independent method even more valuable. We do note that our method results in systematically higher black hole masses than the $M-\sigma$ relation. As we argue in Section~\ref{sec:disc}, this provides a consistent picture on the nature of TDEs in which prompt flares, those that circularized quickly, are expected to be more frequent for higher mass black holes. 

The error bars from \mosfit's measurements of black hole masses in Figure~\ref{fig:Mh_vs_lit} are quite small. Although \mosfit marginalizes over the errors in all of our model's free parameters, it is likely that we are underestimating the total error because our model provides a simple approximation of a complicated physical phenomenon. For example, changing the models for the disrupted stars from ZAMS polytropes with solar composition to more realistic MESA models will prevent the stellar mass of the disrupted star from being uniquely determined without additional knowledge about its evolutionary stage (and through that its radius). This will in turn affect the determination of the peak luminosity and peak timescale, allowing for those parameters to vary more and increasing the uncertainty in the black hole mass. We have accounted for this uncertainty in our systematic errors. Our systematic errors are listed in Table~\ref{table:systematic error} and discussed in the following section.

\begin{figure}[t]
\centering
\includegraphics[width=1 \linewidth,clip=true]{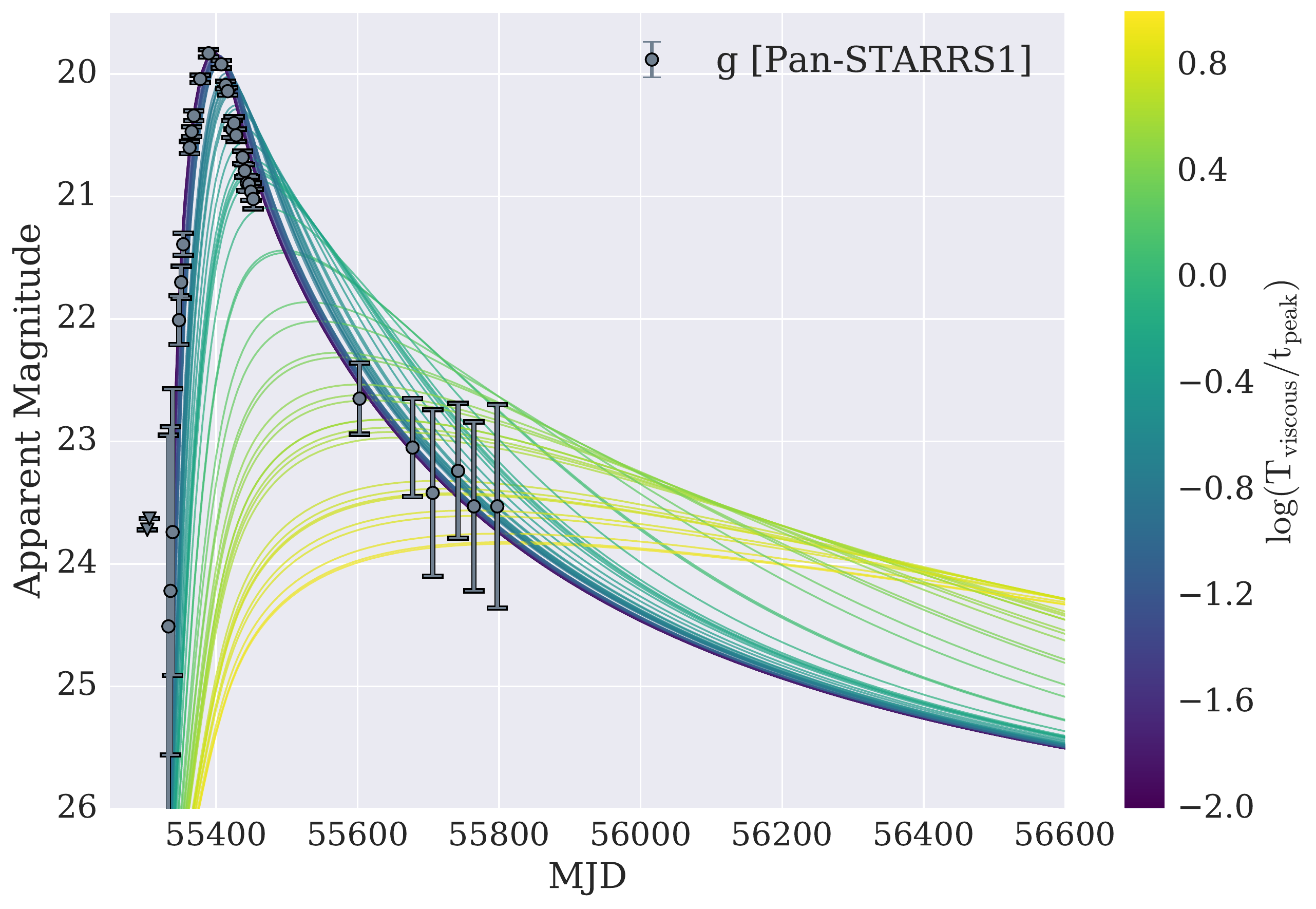} 
\caption{
Example of the effect of a viscous delay on a TDE light curve. The plot shows g-band light curves for PS1-10jh with all parameters but the viscous time set to the best fit values (g-band is shown because it had good coverage over most of the light curve -- all other bands are similarly affected). The best fit light curves are those with no noticeable viscous delays. The plot also shows that $T_{\rm viscous}/t_{\rm peak} \lesssim 0.1$ yields a light curve that is essentially identical to the case with no viscous delay. There were no viscous delays $\gtrsim 10$ days or $\gtrsim 10\%$ of the peak timescale derived in any of the presented fits.} 

\label{fig:varyTvis}
\end{figure}

\subsection{Influence of stellar properties}
The peak timescale of a TDE is primarily determined by the mass of the black hole, and by the mass and radius of the star. As described in Section~\ref{sec:intro}, the effects of the mass and radius for a zero-age, solar metallicity star largely cancel out, allowing the peak timescale to be mostly sensitive to the mass of the black hole. However, varying the age and metallicity of the star can allow the mass and radius to influence the peak timescale to a greater degree. Therefore, the largest systematic uncertainty in our measurements of black hole masses likely comes from uncertainty in the mass-radius relation of the disrupted stars. 

In our model we determined the radius of the stars as a function of their mass. We used the mass-radius relation for ZAMS solar metallicity stars given in \citet{Tout:1996a} for main sequence stars, and set the radii of brown dwarfs to be constant (see Section~\ref{sec:method}). To test how varying the metallicity and age of the stars might affect our measurements, we ran test fits with an additional \textit{radius anomaly} parameter to characterize the uncertainty in the mass-radius relation at each stellar mass. We calculated radius values as a function of mass, metallicity and age for main-sequence stars using \mist \citep{Dotter:2016,Choi:2016,Paxton:2011a,Paxton:2013a,Paxton:2015b}, and used the maximum and minimum radius values at each stellar mass to bound our \textit{radius anomaly} parameter. This was also done for brown dwarfs using the radius values calculated in \citet{Burrows:2011a}. 

We chose a conservative implementation of the \textit{radius anomaly} uncertainty parameter by using a flat prior. Instead of choosing our prior to disfavor unusual age and metallicity combinations, every possible age and metallicity explored in \mist and \citet{Burrows:2011a} was weighted equally. Using the results from these fits we calculated additional systematic errors for each parameter. These error measurements can be found in Table~\ref{table:systematic error}. In general they are significantly larger than the statistical errors quoted in Table~\ref{table:3sigma parameters}; for example, the systematic error in the black hole mass was found to be $\sim 0.2$ dex whereas statistical errors in black hole mass are typically $\lesssim 0.1$ dex.

To further test how changing the mass of the star changes the resulting fit, we performed fits of PS1-10jh while keeping the parameter for the mass of the star constant. We performed these tests for three different star masses: 0.1, 1, and 10 $M_{\odot}$. We found that all three tests achieved comparably good scores, implying that the mass of the star is a degenerate parameter that is difficult to measure accurately with our current model. However, the mass of the black hole does not change dramatically when fixing the stellar mass to different values--despite the uncertainty in the mass of the star we are still able to measure the mass of the black hole. The variation in the black hole mass between tests implies larger uncertainty than our fits in which we leave the stellar mass as a free parameter, however the variation is within the systematic errors we quote in Table~\ref{table:systematic error}. Although only slightly favored by the evidence from the light curve fits, lower mass stars are far more common \citep{Kroupa:1993a} and thus it is likely that the lower stellar masses are closer to the true value. The results from these tests are shown in Table~\ref{table:PS1-10jh comparison table} and are described further in Section~\ref{sec:summary}.

We note that we find a slight preference for stellar masses near $0.1 M_\odot$ (7 events prefer stellar masses between 0.09 and 0.2 $M_\odot$), which is near the peak in the initial mass function. In addition to low mass stars being more common, this preference is likely contributed to by the fact that below this mass the radius of the star no longer cancels out the effect of the mass of the star on the time of peak of the light curve (see Equation~\ref{eqn:tpeak}) -- the mass continues to decrease while the radius remains relatively constant as the star transitions into the brown dwarf regime. For simplicity we assumed the radius was constant below $0.1 M_\odot$ in our current model, although in reality it is likely the radius will actually slightly increase below this mass, see \citet{Burrows:2011a}. This changing mass-radius relationship means that the shortest possible peak times are achieved at $M_{\ast} \sim 0.1 M_\odot$, and thus masses near $0.1 M_\odot$ are favored for events in which short peak times are desired.

\section{Black Hole Mass Estimation}\label{sec:bhestimation}

Without directly imaged stellar orbits (e.g. Sagittarius $\rm A^\ast$), it is very difficult to directly measure black hole masses, and therefore most estimations in the literature are derived using relations between a galaxy's large-scale properties and the size of the black hole at its center. The $M_{\rm h}-\sigma$ relation and the $M_{\rm h}-L_{\rm bulge}$ relation have proven instrumental to our understanding of black holes as a population, but both relations suffer from significant uncertainties. 

\begin{figure}[t!]
\centering
\includegraphics[width=1 \linewidth,clip=true]{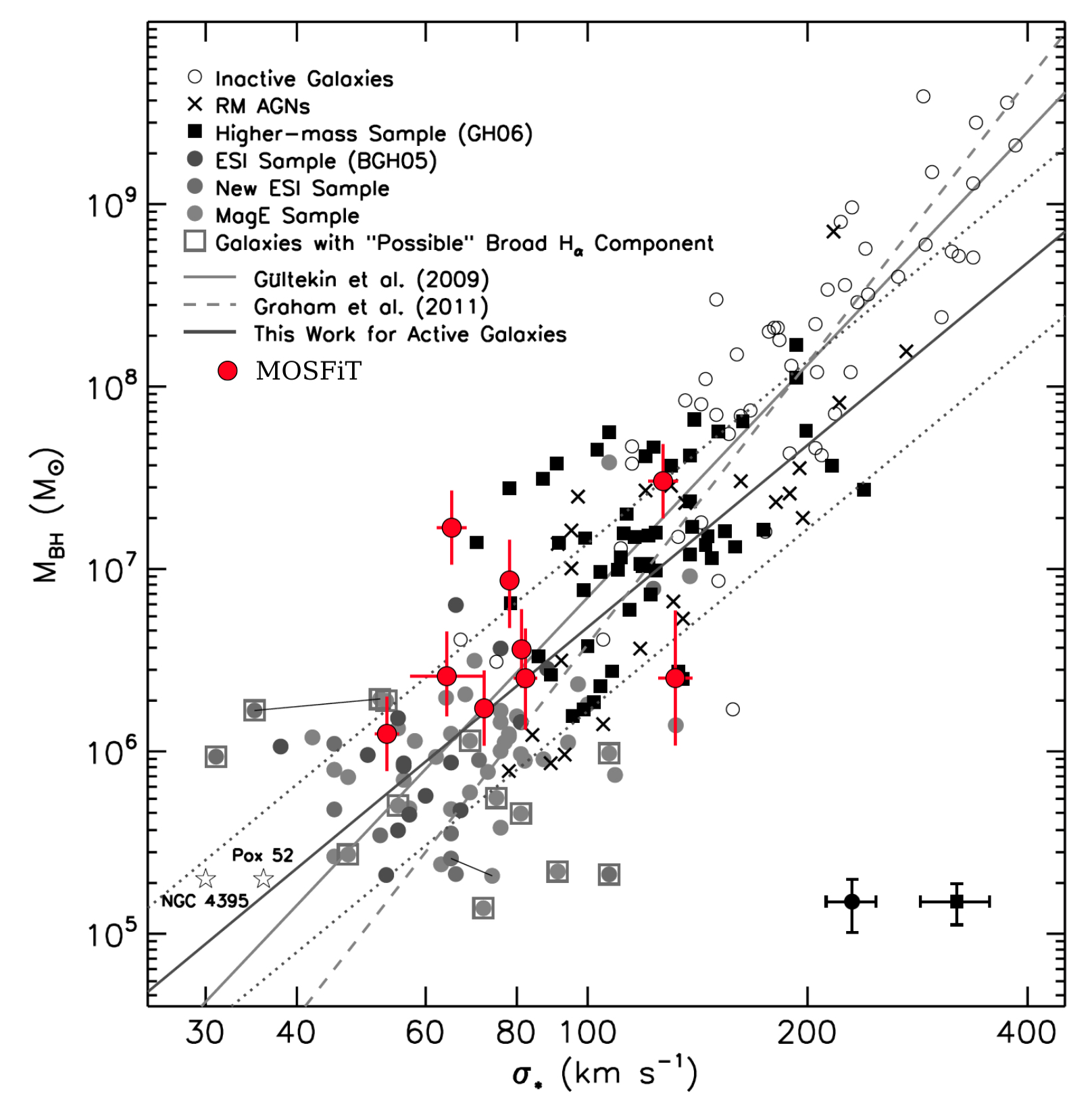} 
\caption{ $M_{\rm h}-\sigma$ for a variety of black hole mass estimates (adapted from \citealt{Xiao:2011}, see that work for details on the original points plotted). The red points show the mass estimates from this work, where the velocity dispersion measurements for our sample of black holes were accumulated from \citet{Thomas:2013,Wevers:2017a,Blagorodnova:2017a, Gezari:2017}.
}
\label{fig:Msigma}
\end{figure}

The intrinsic scatter in the $M_{\rm h}-\sigma$ varies between $\sim 0.3 - 0.5$ dex (it is $\sim 0.46$ dex for the lower mass galaxies in Figure~\ref{fig:Msigma}), and the scatter in the $M_{\rm h}-L_{\rm bulge}$ relation is similarly $\sim 0.5$ dex \citep{McConnell:2013a}. The $M_{\rm h}-\sigma$ relation also changes with galaxy morphology \citep{Hu:2009a,Gadotti:2009a,Graham:2009,Gultekin:2009a}. For example, the relation is a factor of two different for early-type versus late-type galaxies, and differs by an additional factor of two depending on the central density profiles of the galaxies  \citep{McConnell:2013a}.

The dependence of the $M_{\rm h}-\sigma$ and $M_{\rm h}-L_{\rm bulge}$ relation on black hole mass itself has been explored extensively, suggesting that an evolving relationship with mass is likely necessary to minimize scatter. \citet{McConnell:2013a} find that the $M_{\rm h}-L_{\rm bulge}$ appears to have a shallower slope for black holes below $\sim 10^8 M_{\odot}$, and \citet{Jiang:2011} find that the relation differs by over an order of magnitude for black holes between $10^5 M_{\odot}$ and $10^6 M_{\odot}$.
In a study of megamaser galaxies with $M_{\rm h} < 10^8 M_{\odot}$, \citet{Greene:2010a} find that the $M_{\rm h}-\sigma$ relation for larger elliptical galaxies does not hold for their sample of lower mass, maser galaxies. Unfortunately, very few black hole masses have multiple mass measurements, and those that do don't necessarily agree within their error estimates \citep{Peterson:2015}.

Ultimately, accounting for the aforementioned complications can further minimize the scatter about the best-fitting relationship, but these tuned models still make black hole mass predictions that are no better than a factor of $\sim 2$ at all black hole mass scales. While the black hole mass measurements presented in this paper are not always within a 68\% confidence interval of mass measurements from the $M_{\rm h}-\sigma$ relation and the $M_{\rm h}-L_{\rm bulge}$ relation found in the literature, they fit comfortably within the inherent scatter present in both of these relations. In Figure~\ref{fig:Msigma} we overplot the black hole mass measurements from this work on the $M_{\rm h}-\sigma$ relation plot found in \citet{Xiao:2011}, one of the few studies that include a significant number of black hole mass measurements below $10^7 M_{\odot}$. It becomes increasingly difficult to measure black hole mass through direct measurement methods as the mass of the black hole decreases. Most direct measurements of black holes in this mass range, such as the ones in Figure~\ref{fig:Msigma}, come from AGN, as reverberation mapping does not require that the sphere of influence be resolved. 
 
\section{Discussion}\label{sec:disc}

\subsection{Luminosity Follows Fallback Rate}\label{sec:lum follows fallback}
In Section~\ref{sec:results} we briefly discussed how the luminosity appears to closely follow the fallback rate and that none of the events necessitate a viscous delay. Figure~\ref{fig:varyTvis} shows how varying the viscous timescale changes the light curve of PS1-10jh -- it is clear that the data is best fit when $T_{\rm viscous}$ is a very small fraction of $t_{\rm peak}$.

For the luminosity to follow the fallback rate, the debris from the disruption must circularize promptly (or more precisely, while maintaining its initial mass-energy distribution) upon its return to pericenter \citep{Guillochon:2014a}. General relativistic effects are expected to play an important role for disruptions in which $R_{\rm p}$ is comparable to the gravitational radius $R_{\rm g} \equiv G M_{\rm h} / c^2$. Rapid circularization might be achieved through the effects of general relativity, which can strongly influence the trajectories of infalling material. GR precession effects can, for example, cause the stream of infalling debris to intersect itself \citep[e.g.,][]{Dai:2013a}, enabling a dissipation of kinetic energy, as seen in several recent hydrodynamical simulations \citep{Hayasaki:2013a}. This will naturally lead to rapid circularization. 

If spin is included in the calculation, the stream deflects not only within its own orbital plane, but also out of this plane. The result is that the stream does not initially collide with itself \citep{Stone:2012a} and circularization does not immediately occur. If dissipation is minimal, the stream is extremely thin \citep{Kochanek:1994a,Guillochon:2014a} and wraps around the black hole many times \citep{Guillochon:2015a}. In the case of inefficient cooling \citep{Bonnerot:2016b}, the stream can thicken over only a few passages around pericenter, and will intersect with itself more quickly. After a critical number of orbits, stream-stream interactions finally begin to liberate small amounts of gas. This eventually leads to a catastrophic runaway in which all streams simultaneously collapse onto the black hole, circularizing rapidly. For these events, the luminosity should still follow the original fallback rate so long as the mass-energy distribution of the debris remains unchanged (similarly to if rapid circularization had occurred), albeit after a fixed delay time post-disruption. Additionally, once circularization occurs the infalling material is likely to collect around the SMBH into a quasi-spherical layer. This layer is expected to quickly engulf the forming accretion disk, potentially leading to significant reprocessing of the emanated radiation.
\begin{figure}[h!]
\centering
\includegraphics[width=1 \linewidth,clip=true]{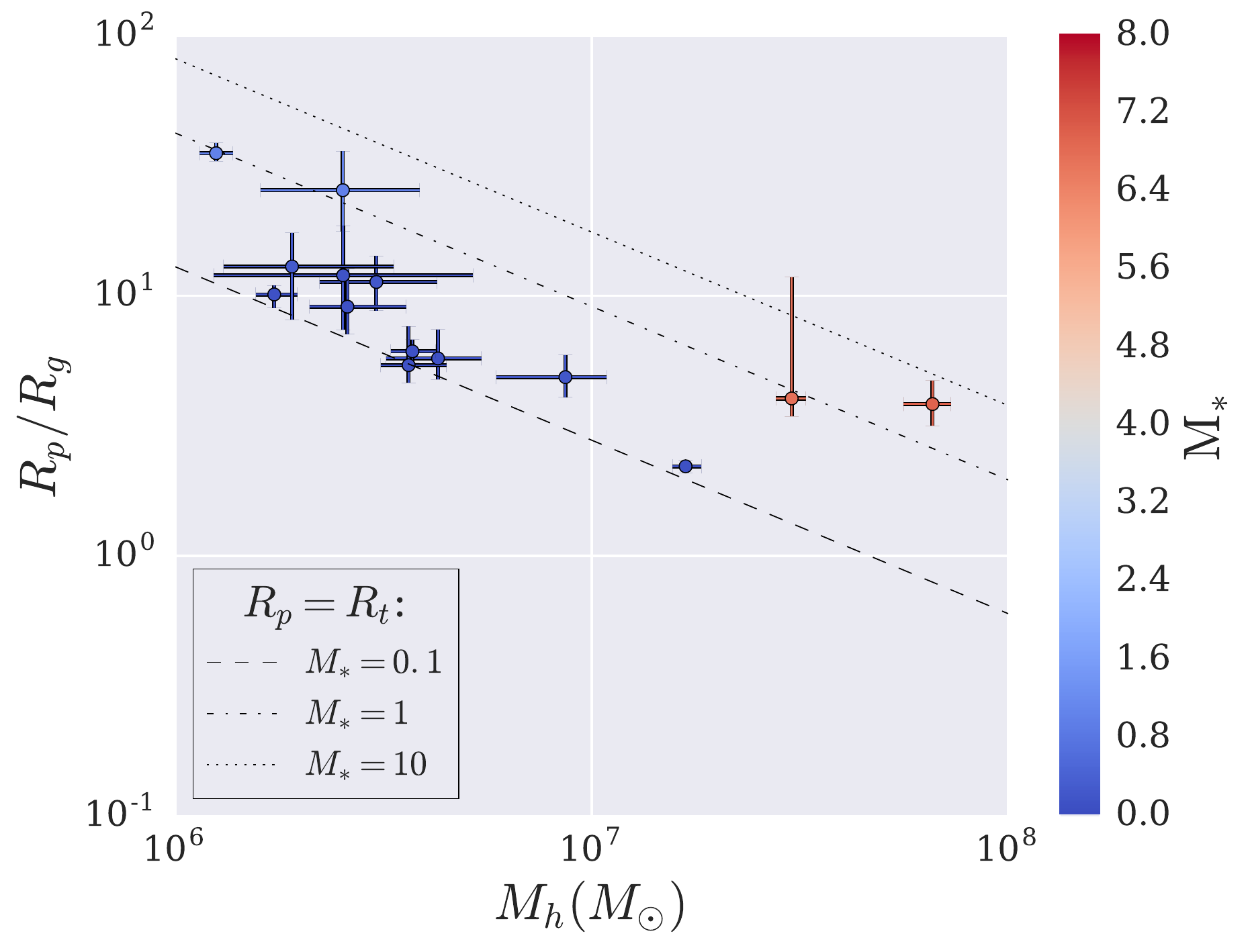} 
\caption{
The dashed lines show $R_{\rm t}/R_{\rm g}$ as a function of $M_{\rm h}$ for differing $M_{\ast}$. Because $R_{\rm t}/R_{\rm g} \propto  {M_h^{-2/3}}$, we expect that lines with slopes of $-2/3$ will map to stars of different masses. Here we have assumed the \citet{Tout:1996a} relations for $R_{\ast}(M_{\ast})$. There is a dependence on the impact parameter as well, and here we have set $\beta = 1$ for the dashed lines, however most of the fits prefer $\beta$ near 1 and, as the plot implies, they also prefer stars between 0.1 and 1 $M_{\odot}$. 
}
\label{fig: Rp/Rs vs Mh}
\end{figure}

In Figure~\ref{fig: Rp/Rs vs Mh} we see that the majority of the fits prefer highly relativistic encounters, which naturally leads to the luminosity following the fallback rate. As mentioned in the previous section, we also find slightly larger black hole masses than those derived using standard galaxy scalings. Larger black holes have larger $R_{\rm g}$ and can thus more easily cause relativistic disruptions. In Figure~\ref{fig: Rp/Rs vs Mh} we show that once $M_{\rm h}$ is a few times $10^7 M_{\odot}$, $R_{\rm g} \approx R_{\rm t}$ for $M_{\ast} \approx 0.1 M_{\odot}$ (the peak of the IMF), meaning that all disruptions in that parameter space are highly relativistic. In general, most of the fits prefer $R_{\rm p}/ R_{\rm g} \lesssim 10$. If $R_{\rm p}/ R_{\rm g}$ is calculated using the black hole masses from the $M-\sigma$ relation (the masses that are systematically smaller than what \mosfit measures), $R_{\rm p}/ R_{\rm g}$ increases from an average value of $\approx 12$ to $\approx 25$ for those disruptions (not all events in this selection have $M- \sigma$ measurements for their black holes).

It has previously been postulated that we should expect a large number of TDEs to be viscously delayed, around 75\% for the black hole mass range probed by the TDEs in this paper \citep{Guillochon:2015a}. Our results imply that we are therefore missing a number of viscously delayed TDEs. It is natural to ask why we seem to be biased towards these prompt, relativistic events. The most obvious explanation is simply that events that fall into this category tend to be easier to detect, as viscous delays can drastically flatten the peak of the light curve, as shown in Figure~\ref{fig:varyTvis}. 

\subsection{Dynamic Reprocessing Layer}

TDEs can result in highly super-Eddington mass fallback rates \citep{De-Colle:2012b}, and therefore we expect excess debris surrounding the black hole to reprocess light from the various dissipation regions \citep{Loeb:1997a,Ulmer:1998a}.

\begin{figure}[h!]
\centering
\includegraphics[width=1 \linewidth,clip=true]{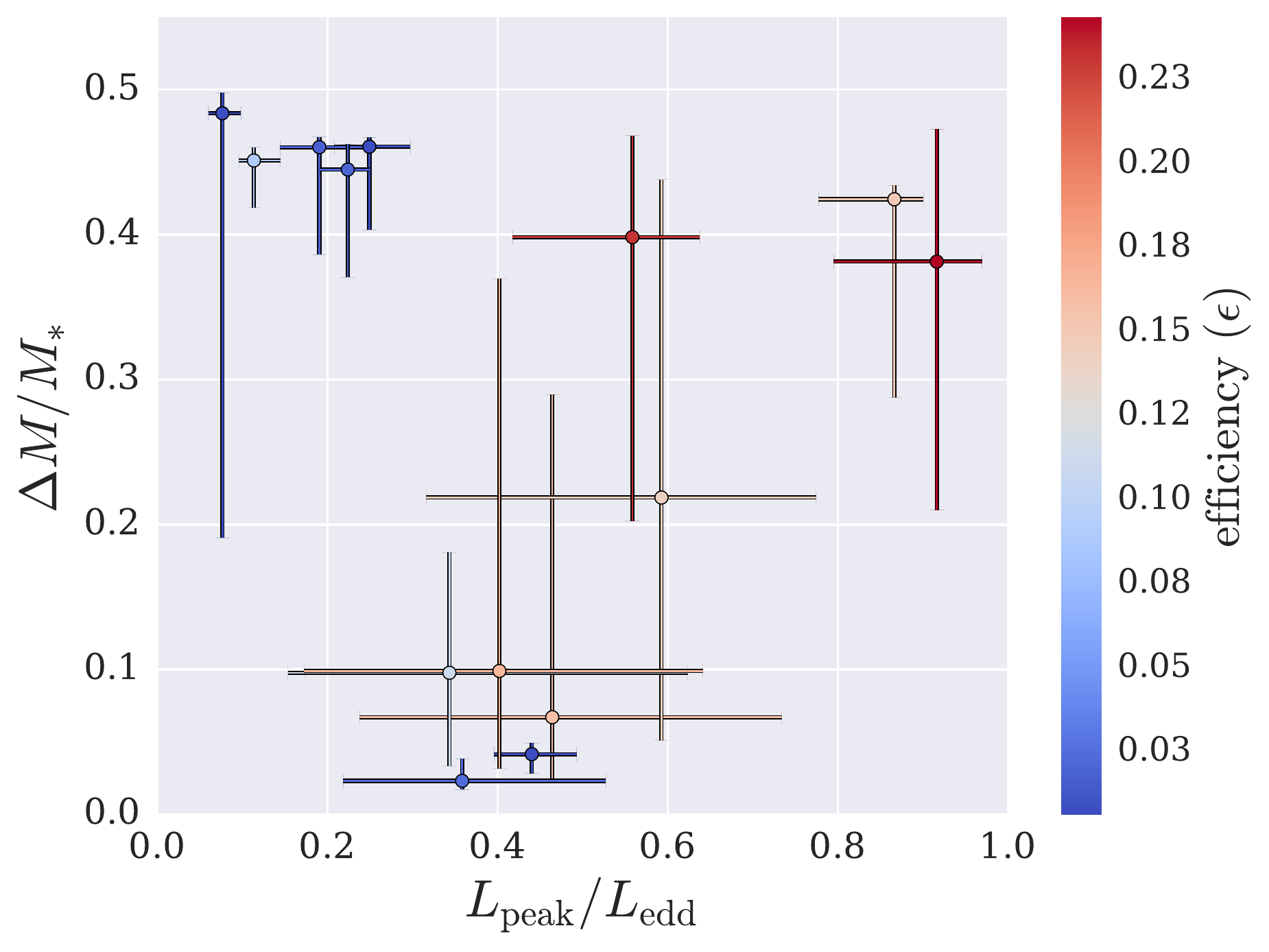} 
\caption{
Fraction of the total stellar mass that remains bound to the black hole versus the fraction of the Eddington limit the peak luminosity reaches.
}
\label{fig:dM/Mstar vs Lpeak/Ledd}
\end{figure}

This is particularly true for the events discussed in this work, as most of them are near full disruption ($\beta_{\rm fd} = 1.8$ for 4/3 polytropes and $\beta_{\rm fd} = 0.9$  for 5/3 polytropes), with large fractions of the disrupted star remaining bound to the black hole, as shown in Figure~\ref{fig:dM/Mstar vs Lpeak/Ledd}.

As our model caps the luminosity of each flare to be no greater than the Eddington limit, our maximum radiated luminosities do not exceed Eddington for any of the modeled flares. However, the fallback rate can exceed the Eddington mass fallback rate (assuming the Eddington mass fallback rate is defined using a constant efficiency, $\dot{M}_{\rm Edd} = \frac{L_{\rm Edd}}{\epsilon c^2} $). In our model, as black holes near their Eddington limit, we implement a soft cut on their luminosity, essentially decreasing the efficiency for the part of the fallback rate that would otherwise result in super-Eddington luminosities. This does not change the average efficiency beyond the quoted errors. However, it does meant that as the luminosity approaches the Eddington luminosity, it becomes more difficult to discern how much mass the black hole is actually accreting as the luminosity depends little on the Eddington excess.

The peak luminosities of most events are $> 10\%$ of their Eddington luminosities, and the peak bolometric luminosity of the fitted events increases with black hole mass, suggesting the luminosities of the events are Eddington limited. Although this runs contrary to the inverse relationship between $L_{\rm peak}$ and $M_{\rm h}$ given by the peak luminosity scaling relation (Equation~(\ref{eqn:lpeak})), this is what we expect for Eddington limited events as $L_{\rm edd} \propto M_{\rm h}$. 

\begin{table}[h!]
\centering
    \renewcommand{\thefootnote}{\arabic{footnote}}
    \footnotesize
    \setlength\tabcolsep{2.pt}
    \renewcommand{\arraystretch}{1.4}
\begin{tabular}{cccc}
\hline
TDE & $T_{\rm phot, \, MOSFiT}$ & $T_{\rm phot, \, lit}$ & source\\
& ($10^3$ K) & ($10^3$ K) & \\
\hline
PS1-10jh & $35_{-2}^{+4}$ & $29_{-2}^{+2}$ & 1, 2 \\
PS1-11af & $22_{-1}^{+1}$ & $19_{-1}^{+1}$ & 3 \\
PTF09djl & $25_{-5}^{+4}$ & $49_{-5}^{+5}$ & 1, 4\\
PTF09ge & $13_{-1}^{+1}$ & $22_{-2}^{+2}$ & 1, 4\\
iPTF16fnl & $33_{-2}^{+2}$ & $35_{-4}^{+4}$ & 1, 5, 6\\
iPTF16axa$^\ast$ & $17_{-2}^{+2}$ & $30_{-3}^{+3}$ & 1, 7\\
ASASSN-14li & $63_{-8}^{+7}$ & $35_{-3}^{+3}$ & 1, 8, 9\\
ASASSN-15oi & $33_{-3}^{+3}$ & $20_{-2}^{+2}$ & 10
\\ 
ASASSN-14ae & $22_{-1}^{+1}$ & $21_{-2}^{+2}$ & 1, 11\\
OGLE16aaa & $23_{-4}^{+6}$ & $> 22$ & 12\\
D1-9$^\ast$ & $110_{-26}^{+33}$ & $55_{-10}^{+10}$ & 13\\
D3-13$^\ast$ & $217_{-6}^{+4}$ & $10_{-1}^{+1}$, $490_{-20}^{+20}$ & 13\\
TDE1 & $34_{-6}^{+7}$ & $29_{-2}^{+2}$ & 1, 14\\
TDE2 & $28_{-3}^{+1}$ & $18_{-1}^{+1}$ & 14\\ 

\hline
\vspace{-1.3em}
\end{tabular}

{
$^1$\citet{Wevers:2017a}, $^2$\citet{Gezari:2012a}, $^3$\citet{Chornock:2014a}, $^4$\citet{Arcavi:2014a}, $^5$\citet{Blagorodnova:2017a}, 
$^6$\citet{Brown:2017a}, $^7$\citet{Hung:2017a}, $^8$\citet{Holoien:2016a}, $^9$\citet{van-Velzen:2016a}, $^{10}$\citet{Holoien:2016b}, $^{11}$\citet{Holoien:2014a},
$^{12}$\citet{Wyrzykowski:2017a},$^{13}$\citet{Gezari:2008a}, $^{13}$\citet{van-Velzen:2011a}

$^*$Temperatures for these events are taken from $\gtrsim 100$ days after the peak of the light curve
}
\caption{
Comparison of photosphere temperatures with literature values. The temperatures in this table were taken near the peak of the light curve with the exception of the three starred ($^*$) TDEs: iPTF16axa, D1-9 and D3-13. The literature values for these events were measured $\gtrsim 100$ days after peak, and so the values quoted from \mosfit were taken as close as possible to the times listed in the source papers for those events. We note that D3-13 has two temperature measurements listed that differ by over an order of magnitude -- this is because \citet{Gezari:2008a} used two blackbodies to fit the optical and UV data for that event. Finally, we note that the value from the literature for ASASSN-14li is dominated by systematic uncertainty not included in the quoted error \citep{Holoien:2016a}.
}
\label{table:Tpeak}
\end{table}

\begin{figure}[h!]
\centering
\includegraphics[width=1 \linewidth,clip=true]{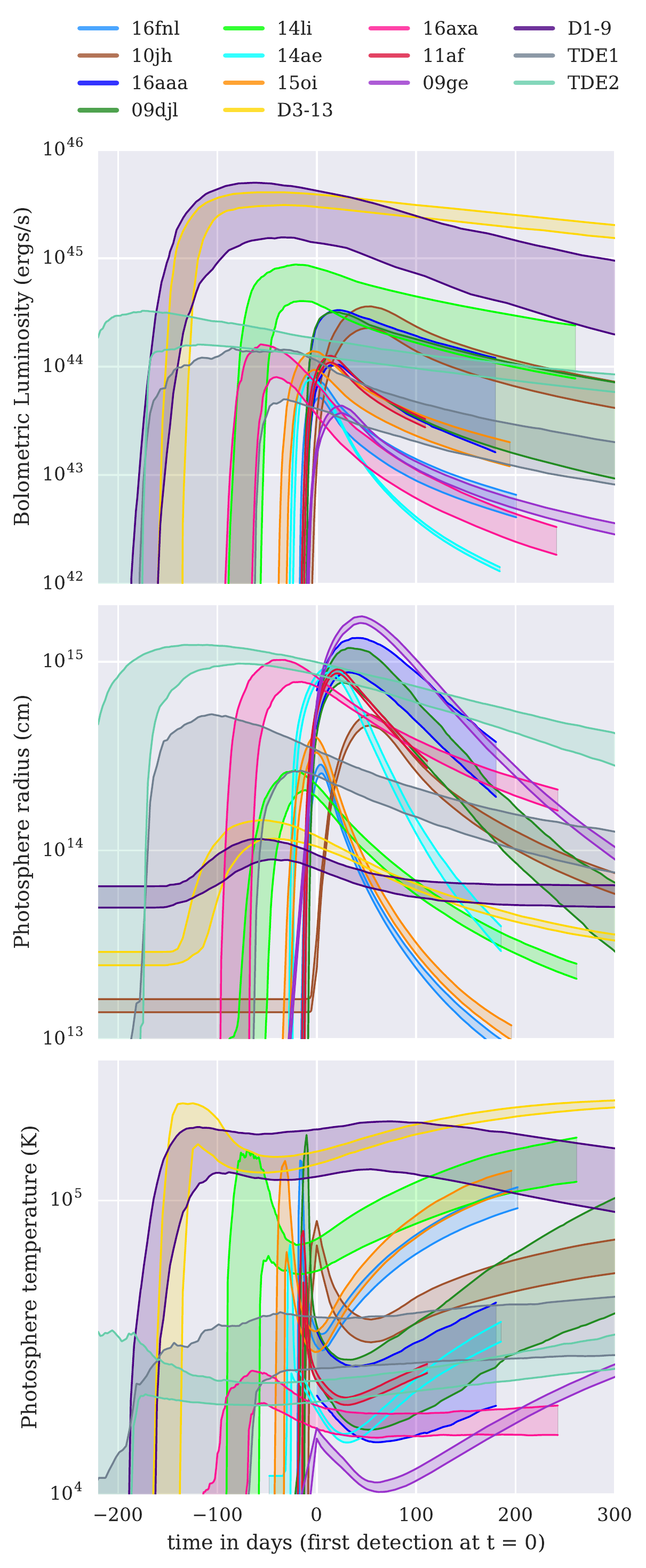}  
\caption{Bolometric luminosity, photosphere radius, and photosphere temperature curves as a function of time since discovery. Each event's curves are colored distinctly and the shaded regions represents the 68\% confidence intervals. The photosphere is approximated as a power law of $L_{\rm bol}$ (see Equation~\ref{eqn:R to L}), and the temperature plotted is the blackbody temperature of the photosphere.}
\label{fig: Lbol, Tphot, Rphot}
\end{figure}

Figure~\ref{fig: Lbol, Tphot, Rphot} shows the relationship between the radius and temperature of this reprocessing layer and the luminosity of the fits. In our fits where we have assumed that the size of the photosphere follows $\dot{M}$ to some power, the temperature we get from the emitting photosphere is comparable with that which has been derived from both fitting blackbodies to the photometry and from spectral observations, with peak values between $2\times 10^4-10^5 K$ (see Figure~\ref{table:Tpeak}).

Although we required the photosphere size to scale as a power law of $\dot{M}$, the parameter range used allowed the exponent of the power law to be zero, which would signify no correlation between $\dot{M}$ and $R_{\rm photo}$. Instead we found that for all fits the exponent was $ > 1/2$ -- the fits required that $R_{\rm photo}$ be a strong function of $\dot{M}$. A similar power law relationship was used to fit the photospheric radius of simulations of TDEs in \citet{Jiang:2016a}, and the power law exponent in that work was found to be $\sim 1$, similar to what we find for some of the event fits presented here.

In Section~\ref{sec:results} we discussed how our model for a growing and shrinking photosphere can help explain additional late time components in TDE light curves. This behavior can also help explain the minimal color evolution present in the light curves. Assuming that the size of the photosphere was set by the tidal radius or the last stable orbit \citep{Loeb:1997a,Ulmer:1998a}, one might expect the temperature to fluctuate as the luminosity varied, as $T \propto L^{1/4}$. However, if the radius of the reprocessing layer increases with luminosity, then $T \propto L^{1/4}/R^{1/2} \propto L^{1/4}/L^{l/2} = L^{1/4 - l/2}$ where $l$ is a power law exponent relating $L$ and $R$ (see Equation~(\ref{eqn:R to L})). As can be seen in Table~\ref{table:3sigma parameters}, we find that most fits prefer $l > 1/2$. Instead of the temperature increasing with luminosity, it decreases slightly near peak and then gradually increases as the luminosity decreases (Figure~\ref{fig: Lbol, Tphot, Rphot}). Because the photosphere temperature is at a local minimum near peak, it can easily match observations that find approximately constant temperature at those times.

This can be interpreted as the result of reprocessing the radiation by a layer of material with optical depth $\tau \sim 1$ in the accretion structures formed by the tidal disruption. The source of this material can be naturally explained by high-entropy material generated by the circularization process, of which only a fraction needs to be ejected to obscure the accretion disk \citep{Guillochon:2014a}. Just as prompt circularization allows the luminosity to follow the fallback rate, it might explain why the reprocessing radius follows the luminosity provided that the obscuring material drains into the black hole on timescales that are short enough to prevent a significant build-up of material. 

Another possible explanation is that the reprocessing layer is generated by a wind or an outflow \citep{Ulmer:1998a,Strubbe:2009a,Miller:2015a,Metzger:2016b}. This is described recently in \citet{Jiang:2016a}, and we find that the temperature evolution seen in Figure~\ref{fig: Lbol, Tphot, Rphot} is reminiscent of the evolution they predict, although the exact power law relations we find between $\dot{M}$ and the photosphere properties show a wider variety of solutions. The \citeauthor{Jiang:2016a} model also predicts temperatures that decrease near peak, because the photospheric radius of the outflow grows with luminosity, and then temperatures that subsequently increase after peak as the ejecta eventually becomes transparent.

\subsection{Summary and Future Prospects}\label{sec:summary}

\begin{itemize}
\item Black hole masses can be accurately measured using tidal disruption events. While the relationship between the time of peak of a TDE and the disrupting black hole's mass was first noted in \citet{Rees:1988a} -- $t_{\rm peak} \propto M_{\rm h}^{1/2}$, it remained unclear until this work if the luminous output of a disruption could be used to measure masses accurately. And although the black hole mass can be estimated from $t_{\rm peak}$ alone, fitting multi-band light curves yields an increased precision of the measurement and makes it possible to learn about other key disruption parameters. Our measurements generally match previous values presented in the literature, as shown in Figure~\ref{fig:Mh_vs_lit}, but we do find some exceptions where the black hole masses acquired from light curve fitting disagree from those derived from galaxy scaling relations.
\item All of the events in this sample have luminosity curves that almost directly follow the fallback of the stellar debris. This requires that the mass-energy distribution remains frozen until it begins to radiate, which can be accomplished through rapid circularization \citep{Hayasaki:2013a,Guillochon:2015b}. When stream intersections occur close to the black hole, the debris is likely to circularize quickly. Because of this, more relativistic encounters with larger impact parameters and black hole masses can increase the likelihood that stream intersections will happen closer to the circularization radius. A lower radiative efficiency in the debris streams can also increase the likelihood that stream intersections occur close to the circularization radius \citep{Bonnerot:2016b}. However, it is unlikely that all TDEs experience rapid circularization \citep{Guillochon:2015b}, and there is still likely to be a class of TDEs that are viscously delayed and are thus generally overlooked in UV/optical surveys.
\item These events are Eddington limited and in most cases significant fractions ($\Delta M/M_{\ast}> 0.1$) of the disrupted stars are bound to the black holes (see Figure~\ref{fig:dM/Mstar vs Lpeak/Ledd}). In these cases there was likely a large amount of stellar debris surrounding the black hole after circularization that could reprocess light from the event.
\item A reprocessing layer that evolves with the bolometric luminosity provides a good match to the optical and UV observations. This could be interpreted as high-entropy material that was generated during the circularization process and then quickly drained into the black hole on timescales short enough to avoid significant build-up. It could also be interpreted as an outflow of material that grows at early times and eventually becomes transparent \citep{Jiang:2016a, Metzger:2016b}. Both of these scenarios could hide the accretion disk from view at early times, preventing X-rays from escaping until the reprocessing layer recedes and/or becomes transparent.
\item Our results suggest that we are (unsurprisingly) biased towards observing the brightest TDEs, which are biased towards the largest black holes when the luminosity is Eddington-limited (but below $\sim 10^8 M_{\odot}$ as most stars are swallowed whole after that point). We find that events in our sample exhibit rapid circularization with no viscous delays lowering the peak luminosity, have luminosities that peak at a significant fraction of their Eddington limits, and are on the high mass end of potential host black holes for tidal disruptions. 
\end{itemize}

While we are able to reliably obtain black hole masses from our analysis of light curves, we find the star and orbit properties are more difficult to determine uniquely. This is likely because the timescale at peak is insensitive to the star's mass, and also because the amount of mass that falls back onto the black hole is degenerate with the efficiency of the radiative process, which we remained agnostic about in this work. As a result, we are often able to find local solutions of similar quality even for radically different efficiency/star mass combinations. While the light curve fits are similar, we suspect that higher efficiency, lower mass solutions are preferable given their improved odds of occurring: low mass stars are significantly more likely to be disrupted than high mass stars. This degeneracy could be broken by a more complete library of stellar disruptions that accounts for relativistic effects \citep[such as black hole spin,][]{Tejeda:2017a} and stellar evolution (which affects composition, rotation, and central concentration) on the debris. Alternatively, the efficiency could be constrained by measuring the properties of the accretion disk and then using limits from these measurements to inform the priors on \mosfit's TDE model. This has been recently attempted for some events, including ASASSN-14li \citep{Cao:2018aa}, PTF09djl \citep{Liu:2017aa}, and PS18kh \citep{Holoien:2018b}. We find that our measured efficiency for ASASSN-14li is significantly higher than the value calculated by \citet{Cao:2018aa} from their modelling of the disk (they find $\epsilon \sim 4\times 10^{-3}$). However, as our systematic uncertainties are large, it is possible the measurements are consistent (\citet{Cao:2018aa} does not quote error values). Our measurements for the efficiency of PTF09djl are consistent with \citet{Liu:2017aa}.
By determining the stellar properties uniquely (or constraining their range by breaking their degeneracy with the radiative efficiency), we could reduce our systematic error in our black hole mass estimates from $\sim 0.2$~dex, to the statistical error bars of an individual model, $\sim 0.1$~dex.

Our current model provides a solid basis for understanding events that radiate most of their energy in the optical/UV. In the future we plan to add an accretion disk component to our model, which will enable fits of TDEs that emit in the X-ray. We also plan to transition to a more realistic library of tidal disruption simulations (e.g. Law-Smith et al. in prep) that utilize MESA models of stars to account for their evolution. As explained above, we expect that this will break the current degeneracy between the mass of the star and the efficiency parameter and allow us to further refine our black hole mass estimates.

\acknowledgments
We thank the anonymous referee for constructive comments and suggestions that helped to improve the manuscript. We thank Jamie~Law-Smith and Nathaniel~Roth for their insightful input. We also thank Thomas~Holoien, Katie~Auchettl, Nicholas~Stone, Sjoert~van Velzen, Phillip~Macias, Thomas~Wevers, Suvi~Gezari, Benjamin~Shappee, Christopher~Kochanek,  Ilya~Mandel, Kate~Alexander, Iair~Arcavi and Jane~Dai for useful comments and discussions. Additionally, we thank Thomas~Holoien, Jonathan~Brown, and Nadejda~Blagorodnova for supplying host-subtracted data for the ASASSN events and iPTF16fnl. E.R.-R. and B.M  are grateful for support from the Packard Foundation, DNRF, NASA ATP grant NNX14AH37G and NSF grant AST-161588. The calculations for this research were carried out in part on the UCSC supercomputer Hyades, which is supported by National Science Foundation (award number AST-1229745) and UCSC. 

\software{This research has made use of NASA's Astrophysics Data System, {\tt SciPy} \citep{Jones:2001a}, {\tt Astropy} \citep{Astropy-Collaboration:2013a}, {\tt NumPy} \citep{Van-Der-Walt:2011a}, and {\tt emcee} \citep{Foreman-Mackey:2013a}.} 

\bibliographystyle{yahapj}
\bibliography{library}

\appendix\label{sec:appendix}

\begin{table}[ht]
    \renewcommand{\thefootnote}{\arabic{footnote}}
    \footnotesize
    \setlength\tabcolsep{3pt}
    \renewcommand{\arraystretch}{1.4}
\centering
\begin{tabular}{ccccccccccc}
\hline
$M_{\rm h}$ & $\beta$ & $M_{\ast}$ & $\gamma$ &$\epsilon$ & $\rm log(R_{\rm ph0})$ & $l$ & $t_{\rm peak}$ & $T_{vis}$ & WAIC & PSRF \\
($10^6 M_{\odot}$) & & ($M_{\odot}$) & & & & & (days) & (days) & &  \\
\hline
$12.6_{-1.1}^{+0.9}$ & $1.81_{-0.01}^{+0.03}$ & $10.0$ & 4/3 & $0.0004_{-0.0001}^{+0.0001}$ & $0.28_{-0.06}^{+0.06}$ &$0.77_{-0.05}^{+0.05}$ & $74_{-2}^{+2}$ & $0.3_{-0.3}^{+3.5}$ & 199.8 & 1.012 \\
$8.6_{-0.7}^{+0.7}$ & $1.81_{-0.02}^{+0.03}$ & 1.0 & 4/3 & $0.0038_{-0.0005}^{+0.0006}$ & $0.25_{-0.05}^{+0.06}$ & $0.73_{-0.03}^{+0.05}$ & $73_{-2}^{+2}$ & $0.4_{-0.4}^{+4.2}$ & 193.5 & 1.033 \\
$17_{-1}^{+2}$ & $0.90_{-0.01}^{+0.01}$ & 0.1 & 5/3 & $0.09_{-0.02}^{+0.02}$ & $0.8_{-0.1}^{+0.1}$ & $1.44_{-0.09}^{+0.08}$  & $111_{-2}^{+4}$ & $0.05_{-0.05}^{+1.40}$ & 200.4 & 1.034 \\
\hline

\end{tabular}
\caption{Comparison between test runs of PS1-10jh with $M_{\ast}$ parameter set to different constant values: 0.1, 1.0, 10.0 $M_{\odot}$. While all runs converged with similar scores, we expect the run with $M_{\ast} = 0.1 M_{\odot}$ to be the most likely true solution as these stars are much more common and are more likely to be disrupted.}
\label{table:PS1-10jh comparison table}
\end{table}

\begin{table}[h!]
\centering
    \renewcommand{\thefootnote}{\arabic{footnote}}
    \footnotesize
    \setlength\tabcolsep{2.5pt}
    \renewcommand{\arraystretch}{1.4}
\begin{tabular}{cccc}
\hline
TDE & $R_{\rm p}/R_{\rm g}$ & $L_{\rm bol}/L_{\rm edd}$ & $\Delta M/M_{\ast}$ \\
\hline
iPTF16fnl & $10.1_{-1.1}^{+0.8}$ & $0.25_{-0.04}^{+0.05}$ & $0.46_{-0.06}^{+0.01}$ \\
PS1-10jh & $2.2_{-0.1}^{+0.1}$ & $0.11_{-0.02}^{+0.03}$ & $0.45_{-0.03}^{+0.01}$ \\
OGLE16aaa & $11_{-3}^{+3}$ & $0.5_{-0.2}^{+0.3}$ & $0.07_{-0.04}^{+0.22}$ \\
PTF09djl & $9.1_{-2.0}^{+3.6}$ & $0.6_{-0.3}^{+0.2}$ & $0.2_{-0.2}^{+0.2}$ \\
ASASSN-14li & $4.9_{-0.8}^{+1.0}$ & $0.56_{-0.14}^{+0.08}$ & $0.40_{-0.20}^{+0.07}$ \\
ASASSN-14ae & $35_{-2}^{+3}$ & $0.44_{-0.04}^{+0.05}$ & $0.04_{-0.01}^{+0.01}$ \\
ASASSN-15oi & $6_{-1}^{+2}$ & $0.19_{-0.05}^{+0.05}$ & $0.46_{-0.07}^{+0.01}$ \\
D3-13 & $4.0_{-0.6}^{+7.8}$ & $0.87_{-0.09}^{+0.03}$ & $0.42_{-0.14}^{+0.01}$ \\
iPTF16axa & $25_{-8}^{+11}$ & $0.4_{-0.1}^{+0.2}$ & $0.02_{-0.01}^{+0.02}$ \\
PS1-11af & $6.1_{-0.6}^{+0.7}$ & $0.22_{-0.03}^{+0.03}$ & $0.45_{-0.07}^{+0.02}$ \\
PTF09ge & $5.4_{-0.8}^{+2.2}$ & $0.08_{-0.02}^{+0.02}$ & $0.48_{-0.29}^{+0.01}$ \\
D1-9 & $3.8_{-0.7}^{+0.9}$ & $0.3_{-0.2}^{+0.3}$ & $0.10_{-0.06}^{+0.08}$ \\
TDE1 & $12_{-5}^{+7}$ & $0.4_{-0.2}^{+0.2}$ & $0.10_{-0.07}^{+0.27}$ \\
TDE2 & $13_{-5}^{+5}$ & $0.92_{-0.12}^{+0.05}$ & $0.38_{-0.17}^{+0.09}$ \\
\hline
\vspace{-1.3em}
\end{tabular}
\caption{Tabulated values from Figure~\ref{fig: Rp/Rs vs Mh} and Figure~\ref{fig:dM/Mstar vs Lpeak/Ledd}.}
\label{table:tabulated figure values}
\end{table}

\end{document}